\documentclass[12pt]{article}
\usepackage{amssymb,amsmath,epsfig}
\usepackage{graphicx}
\allowdisplaybreaks
\begin{document}

\title{\bf Anisotropic Strange Stars through Embedding Technique in Massive Brans-Dicke Gravity}

\author{M. Sharif \thanks{msharif.math@pu.edu.pk} and Amal Majid
\thanks{amalmajid89@gmail.com}\\
Department of Mathematics, University of the Punjab,\\
Quaid-e-Azam Campus, Lahore-54590, Pakistan.}

\date{}

\maketitle
\begin{abstract}
This paper investigates the existence and properties of anisotropic
strange quark stars in the context of massive Brans-Dicke theory.
The field equations are constructed in Jordan frame by assuming a
suitable potential function with MIT bag model. We employ the
embedding class-one approach as well as junction conditions to
determine the unknown metric functions. Radius of the strange star
candidate, LMC X-$4$, is predicted through its observed mass for
different values of the bag constant. We analyze the effects of
coupling parameter as well as mass of scalar field on state
determinants and execute multiple checks on the stability and
viability of the spherical system. It is concluded that the
resulting stellar structure is physically viable and stable as it
satisfies the energy conditions as well as essential stability
criteria.
\end{abstract}

\section{Introduction}

In the field of astronomy, the observational data of compact stellar
structures provide information regarding their mass, rate of
rotation and emitted radiations. However, in order to understand
their internal mechanism and evolutionary process, we rely on
analytical methods of relativistic theories. Among compact objects,
intriguing nature of neutron stars has motivated researchers to
explore their composition, structure and other features. Neutron
star has a core of $1$ to $3$ solar masses $(M_{\bigodot})$ which
resists further collapse by counterbalancing the inward pull of
gravity through degeneracy pressure of newly generated neutrons.
These celestial bodies either exist individually or have a companion
to form binary systems. The discovery of neutrons led to the
prediction of neutron stars in $1934$ \cite{1} but observational
evidence came later. This is because neutron stars do not emit
enough radiation and are mostly undetectable. Generally, they are
spotted as rapidly rotating pulsars which emit radiation at regular
intervals ranging from milliseconds to seconds. The first pulsar was
discovered in $1967$ pulsating for $0.3$ seconds after every $1.37$
seconds \cite{2}. Some stellar candidates of pulsars include 4U
1820-30, Her X-1, PSR J1903+327, etc.

The study of relativistic objects has revealed that physical
properties vary with changes in direction, i.e., they are
anisotropic in nature. Anisotropy may occur in both low and high
density profiles due to a large number of physical processes such as
rotational motion, phase transition or presence of magnetic field or
viscous fluid. Ruderman \cite{2'} proposed that highly dense system
of interacting nuclear matter introduces anisotropy. Since dense
cores of stellar objects exhibit extreme nuclear density, anisotropy
is one of the dominant features of their intrinsic geometry and
evolution. Researchers have explored the effects of anisotropy by
considering transverse and radial components of pressure. Herrera
and Santos \cite{3} discussed the plausible causes and effects of
local anisotropy in self-gravitating systems. Harko and Mak \cite{4}
found static interior solutions for anisotropic relativistic objects
by considering a specific anisotropy factor. The stability of
anisotropic structures including the effects of cosmological
constant was explored by Hossein et al. \cite{5}. Paul and Deb
\cite{6} formulated physically viable solutions for anisotropic
compact stars in hydrostatic equilibrium.

Apart from the three outcomes of collapse (white dwarf, neutron star
and black hole), another compact stellar structure is also
hypothesized as an end state of inward fall of a neutron star. It is
believed that such a cosmic object is composed of strange quark
matter which is a favorable state of baryon matter. Witten \cite{8}
identified two possibilities of formation of quark matter: the
quark-hadron phase transition in the early universe or the
transformation of neutron stars into quark stars at extremely high
densities. A strange quark star is an intermediate stage between
black hole and neutron star which has too much mass at its core for
the neutrons to hold their individuality but still evades collapse
into a black hole. Recent observational estimates of masses and
radii of some stars (Her X-$1$, 4U 1820-30, LMC X-4, etc.) are not
consistent with neutron star prototype. Instead they can be treated
as suitable candidates for strange quark stars.

The equation of state (EoS) of compact objects (such as neutron and
quark stars) is not yet determined despite the existence of several
models. The quark star is composed of stable strange quark matter
(SQM) (general Witten's conjecture \cite{8}) made of equal number of
up, down and strange quarks and is assumed to be the true ground
state for the confined hadrons \cite{101}. Interestingly, the
neutron star EoS failed to explain the compactness of the compact
stellar objects like 4U 1820-30, SAX J 1808.4-3658, 4U 1728-34, Her
X-1, RXJ 185635-3754 and PSR 0943+10, etc., whereas SQM EoS (MIT Bag
model) \cite{102} has satisfactorily explained the compactness of
the stellar candidates. Recent observations of gravitational waves
from binary neutron stars collision (GW170817 \cite{104} and
GW190425 \cite{105}), have made it possible to estimate the range of
masses and thus constraint the mass of neutron and quark stars,
determining the MIT bag model as the best approximation for the EoS.

The bag constant ($\mathcal{B}$) appearing in the EoS evaluates the
difference between energy density of true (global minimum of energy
with stable configuration) and false (local minimum of energy with
unstable configuration) vacuum. Increasing the bag constant lowers
the quark pressure ultimately affecting the stellar structure. Many
people \cite{9} have considered the MIT bag model as an EoS for
predicting the interior distribution of quarks in strange stars.
Rahaman et al. \cite{10} developed a new interpolating function for
calculating the mass of strange stars and explored physical features
of a star of radius $9.9km$. A model for a hybrid star composed of
normal as well as quark matter was presented by Bhar \cite{11} in
the context of Krori and Barura ansatz. Arba\~{n}il and Malheiro
\cite{12'} analyzed the impact of anisotropy on stability as well as
equilibrium of quark stars through equations of radial oscillation
and hydrostatic equilibrium. The effect of electromagnetic field on
anisotropic strange star models has been investigated by employing
MIT bag model \cite{12}. Deb et al. \cite{13} studied singularity
free solutions representing uncharged as well as charged quark stars
with the help of MIT bag model and checked their viability as well
as stability. Bhar \cite{15} constructed anisotropic model for
strange stars using the condition for embedding class-one and
checked it for viability and stability.

General relativity (GR) has been accepted as the best fit for
describing relativistic structures but suffers from setbacks when
defining the current phase of the universe. A natural extension of
GR is obtained by including a scalar field that mediates gravity
over long ranges. These theories are known as scalar-tensor theories
\cite{19} and have extensively been used for discussing different
astronomical phenomena. Brans-Dicke (BD) theory, a prototype of
scalar-tensor theory, satisfies Mach's principle as well as couples
a massless scalar field $\bar{\Phi}$ and metric tensor
$g_{\gamma\delta}$ to matter field through a coupling constant
$\omega_{BD}$ \cite{20}. In the background of BD theory,
$\omega_{BD}$ acts like a tuneable parameter that can take on
different values to acquire desirable results.

The value of $\omega_{BD}$ must exceed the value $40,000$ in order
to satisfy weak field experiments \cite{21} whereas the inflationary
model holds for lower values of $\omega_{BD}$ \cite{22}. The
conflict is resolved by replacing the massless scalar field with a
massive scalar field $\Phi$ and adding a self-interacting potential
function $V(\Phi)$ in BD theory leading to massive BD (MBD) gravity.
The scalar field mass ($m_\Phi$) leads to a finite range of the
scalar field of the order of its compton wavelength
($\lambda_\Phi$). For $m_\Phi\gtrsim2\times10^{-25}GeV$ (or
$\lambda_\Phi\lesssim10^{11}m$), the solar system observations
cannot put stringent restrictions on the BD parameter and all values
of $\omega_{BD}$ greater than $-\frac{3}{2}$ are allowed \cite{23}.
Like all scalar-tensor theories, MBD can be discussed in both Jordan
and Einstein frames based on the nature of coupling (minimal or
non-minimal) of scalar field to matter. In Einstein frame, the
coupling function $\alpha(\Phi)$ is related to coupling parameter
$\omega_{BD}$ in Jordan frame as $\alpha^2=(2\omega_{BD}+3)^{-1}$.

The slowly and rapidly rotating neutron stars have extensively been
discussed in scalar-tensor theories. Sotani \cite{24} studied
neutron stars in a massless scalar field and concluded that
deviations from GR increase with the increase in mass of celestial
objects. Silva et al. \cite{25} discussed the effect of anisotropy
on moment of inertia of rotating neutron stars. The structure and
properties of slowly rotating neutron stars have also been explored
under the influence of a massive scalar field. Doneva and Yazadjiev
\cite{26} investigated the dynamics of rapidly rotating neutron
stars in the presence of a massive scalar field and concluded that
deviations from GR can be large due to larger moment of inertia.
Staykov et al. \cite{27} extended this work by considering a
self-interacting potential along with a massive scalar field to
analyze the behavior of static and slowly rotating neutron stars.
Salient features as well as validity of different models for strange
quark stars have also been examined in modified theories like $f(R)$
and $f(R,T)$ gravity \cite{28, 28a}. Recently, we have studied
self-gravitating systems by deriving anisotropic extensions of
isotropic solutions through gravitational decoupling technique in
the context of MBD theory \cite{27'}.

In this paper, we explore physical attributes of strange stars and
discuss their existence in the context of MBD theory. The paper is
organized as follows. In section \textbf{2}, we construct a system
of field equations and physical variables using MIT bag model.
Section \textbf{3} gives an overview of junction conditions for a
smooth matching between intrinsic and extrinsic geometries. The
physical properties, validity and stability are examined in section
\textbf{4}. In the last section, we summarize our results.

\section{Massive Brans-Dicke Theory and Matter Variables}

Scalar-tensor theories can be represented in Jordan as well as
Einstein frames, which are related through a conformal
transformation. The action of scalar-tensor theories in Jordan frame
\cite{30} with $G_0=1$ is defined as
\begin{equation}\label{0}
S=\int\sqrt{-g}(\mathcal{R}\Phi-\frac{\omega_{BD}}{\Phi}\nabla^{\gamma}\nabla_{\gamma}\Phi
-V(\Phi)+\emph{L}_m)d^{4}x,
\end{equation}
where $g,~\mathcal{R}$ and $L_m$ represent determinant of the metric
tensor, Ricci scalar and matter lagrangian, respectively. The
function $V(\Phi)$ completely specifies the scalar-tensor theory.
For the present study, we choose
\begin{equation}\label{0'}
V(\Phi)=\frac{1}{2}m_{\Phi}^2\Phi^2.
\end{equation}
Both slowly and rapidly rotating neutron stars have already been
studied for this form of potential function \cite{26, 26a}. The
metric $\hat{g}_{\gamma\delta}$ and scalar field $\hat{\Phi}$ can be
obtained for Einstein frame through the transformations
$\hat{g}_{\gamma\delta}=\mathcal{A}^{-2}(\Phi){g}_{\gamma\delta}$
and $\Phi=\mathcal{A}^{-2}(\hat{\Phi})$. The variation of action
(\ref{0}) with respect to $g_{\gamma\delta}$ and $\Phi$ yields the
field equations and evolution equation, respectively, given as
\begin{eqnarray}\nonumber
G_{\gamma\delta}&=&\frac{1}{\Phi}[T_{\gamma\delta}^{(m)}+T_{\gamma\delta}^{\Phi}]=\frac{1}{\Phi}[T_{\gamma\delta}^{(m)}
+\Phi_{,\gamma;\delta}-g_{\gamma\delta}\Box\Phi+\frac{\omega_{BD}}{\Phi}
(\Phi_{,\gamma}\Phi_{,\delta}\\\label{1}
&-&\frac{g_{\gamma\delta}\Phi_{,\mu}\Phi^{,\mu}}{2})-\frac{V(\Phi)g_{\gamma\delta}}{2}],\\\label{2}
\Box\Phi&=&\frac{T^{(m)}}{3+2\omega_{BD}}+\frac{1}{3+2\omega_{BD}}
(\Phi\frac{dV(\Phi)}{d\Phi}-2V(\Phi)),
\end{eqnarray}
where the energy-momentum tensor $T_{\gamma\delta}^{(m)}$ represents
the matter distribution and $T^{(m)}$ is its trace with $\Box$ being
the d'Alembertian operator.

We assume that the static spherical structure of the stellar object
is described by the line element
\begin{equation}\label{3}
ds^2=e^{\nu(r)}dt^2-e^{\lambda(r)}dr^2-r^2(d\theta^2+\sin^2\theta
d\phi^2),
\end{equation}
where $\nu(r)$ and $\lambda(r)$ are metric potentials. Stellar
systems are characterized by anisotropic pressure and inhomogeneous
energy density which take a dominant part in their evolution. For
this purpose, we discuss the physical features of strange stars with
anisotropic distribution specified by the following energy-momentum
tensor
\begin{equation}\label{4}
T_{\gamma\delta}^{(m)}=(\rho+p_\perp) u_{\gamma}u_{\delta}-p_\perp
g_{\gamma\delta}+(p_r-p_\perp)s_{\gamma}s_{\delta},
\end{equation}
where $u_\gamma=(e^{\frac{\nu}{2}},0,0,0)$ is the 4-velocity of
comoving observer and $s_{\gamma}=(0,-e^{\frac{\lambda}{2}},0,0)$ is
a radial 4-vector. Here, $\rho,~p_r$ and $p_\perp$ represent the
energy density, radial and transverse pressures, respectively. Using
Eqs.(\ref{1})-(\ref{4}), the field equations are obtained as
\begin{eqnarray}\label{5}
\frac{1}{r^2}-e^{-\lambda}\left(\frac{1}{r^2}-\frac{\lambda'}{r}\right)&=&
\frac{1}{\Phi}({\rho}+T_0^{0\Phi}),\\\label{6}
-\frac{1}{r^2}+e^{-\lambda}\left(\frac{1}{r^2}+\frac{\nu'}{r}\right)&=&
\frac{1}{\Phi}({p_r}-T_1^{1\Phi}),\\\label{7}
\frac{e^{-\lambda}}{4}\left(2\nu''+\nu'^2-\lambda'\nu'+2\frac{\nu'-\lambda'}{r}\right)&=&
\frac{1}{\Phi}({p}_{\perp}-T_2^{2\Phi}),
\end{eqnarray}
where prime denotes differentiation with respect to $r$ and the
expressions of $T_0^{0\Phi},~T_1^{1\Phi}$ and $T_2^{2\Phi}$ are
given in Appendix A. The wave equation (\ref{2}) turns out to be
\begin{eqnarray}\nonumber
\Box\Phi&=&-e^{-\lambda}\left[\left(\frac{2}{r}-\frac{\lambda'} {2}
+\frac{\nu'}{2}\right)\Phi'(r)+\Phi''(r)\right],\\\label{8}
&=&\frac{1}{3+2\omega_{BD}}\left[T^{(m)}+
\left(\Phi\frac{dV(\Phi)}{d\Phi}-2V(\Phi)\right)\right].
\end{eqnarray}

It has been shown that if a symmetric tensor $b_{\gamma\delta}$
satisfies the Gauss-Codazi equations given as
\begin{equation}\label{10}
R_{\gamma\delta\mu\upsilon}=2eb_{\gamma[\mu}b_{\upsilon]\delta}\quad\text{and}\quad
b_{\gamma[\delta;\mu]}-\Gamma^{\lambda}_{\delta\mu}b_{\gamma\lambda}
+\Gamma^\lambda_{\gamma[\delta}b_{\mu]\lambda}=0,
\end{equation}
the $(n+1)$ dimensional space can be embedded in an $(n+2)$
dimensional pseudo-Euclidean space \cite{31}. Here $e=\pm1$,
$R_{\gamma\delta\mu\upsilon}$ denotes curvature tensor and
$b_{\gamma\delta}$ are the co-efficients of second differential
form. From the above equation, Eiesland \cite{32} obtained a
necessary and sufficient condition for an embedding class-one as
\begin{equation}\label{10'}
R_{0101}R_{2323}-R_{1212}R_{0303}-R_{1202}R_{1303} =0,
\end{equation}
which leads to the following differential equation for the
considered metric
\begin{equation}\label{10*}
(\lambda'-\nu')\nu'e^\lambda+2(1-e^\lambda)\nu''+\nu'^2=0.
\end{equation}
The solution of the above equation turns out to be
\begin{equation}\label{12}
\lambda(r)=\ln(1+B\nu'^2e^\nu),
\end{equation}
where $B$ is a constant of integration. Maurya et al. \cite{33}
constructed a new class of solutions using the following form of
metric potential
\begin{equation}\label{11}
\nu(r)=2r^2A+\ln C,
\end{equation}
where $A$ and $C$ are positive constants. Using this value in
Eq.(\ref{12}), we have
\begin{equation}\label{11'}
\lambda(r)=\ln(1+ADr^2e^{2Ar^2}),
\end{equation}
where $D=16ABC$ is a constant.

Neutron stars with $M>3M_{\bigodot}$ may transform into quark stars
which contain up $(u)$, down $(d)$ and strange $(s)$ quark flavors.
The matter variables describing the interior configuration of these
relativistic stars obey MIT bag EoS. According to the MIT bag model,
the quark pressure is stated as
\begin{equation}\label{7}
p_r=\sum_{f}p^f-\mathcal{B},\quad f=u,~d,~s,
\end{equation}
where $p^f$ corresponds to the individual pressure of each quark
flavor which is neutralized by the total external bag pressure
$\mathcal{B}$, also known as bag constant. The total energy density
of deconfined quarks is defined by the bag model as
\begin{equation}\label{8}
\rho=\sum_{f}\rho^f+\mathcal{B},
\end{equation}
where energy density of each flavor $\rho^f$ is related to the
respective pressure as $\rho^f=3p^f$. The EoS of MIT bag model for
strange stars is inferred from Eqs.(\ref{7}) and (\ref{8}) as
\begin{equation}\label{9}
p_r=\frac{1}{3}(\rho-4\mathcal{B}).
\end{equation}
The simplified form of this EoS has been used in GR and modified
theories to examine the features of quark star candidates. In our
study, the numerical results of the model have been obtained by
taking $\mathcal{B}$ equal to $64MeV/fm^3$ and $83MeV/fm^3$ which
are within the allowed limit \cite{10}. The total mass of a sphere
of radius $r$ is evaluated through Misner-Sharp formula as
\begin{equation}\label{9'}
m=\frac{r}{2}(1-e^{-\lambda}).
\end{equation}

\subsection{Matching Conditions}

The set of parameters ($A,~B,~C,~D$) defining the geometry as well
as physical properties (such as mass and radius) of anisotropic
compact objects can be determined through the smooth matching of
interior and exterior spacetimes on the boundary ($\Sigma$) of the
star. The exterior region is taken to be the Schwarzschild spacetime
given by
\begin{equation}\label{49}
ds^2=(1-\frac{2M}{r})dt^2-\frac{1}{(1-\frac{2M}{r})}dr^2
-r^2(d\theta^2+\sin^2\theta d\phi^2),
\end{equation}
where $M$ is the mass. To ensure smoothness and continuity of
geometry at the boundary surface, the following conditions must be
satisfied at the hypersurface $\Sigma$ $(f=r-R=0, R$ is constant
radius)
\begin{eqnarray}\label{13}
&(ds^2_{-})_{\Sigma}=(ds^2_{+})_{\Sigma},\quad
(K_{ij_-})_{\Sigma}=(K_{ij_+})_{\Sigma},&\\
&(\Phi(r)_-)_{\Sigma}=(\Phi(r)_+)_{\Sigma},\quad
(\Phi'(r)_-)_{\Sigma}=(\Phi'(r)_+)_{\Sigma}.&
\end{eqnarray}
Here $K_{ij}$ denotes curvature whereas subscripts $-$ and $+$
represent interior and exterior spacetimes, respectively. The
continuity of the the first fundamental form ($[ds^2]_\Sigma=0$)
leads to
\begin{equation*}
[H]_\Sigma\equiv H(r\rightarrow R^+)-H(r\rightarrow R^-)\equiv
H^+_R-H_R^-,
\end{equation*}
for any function $H(r)$. The above condition yields
$g_{tt}^-(R)=g_{tt}^+(R)$ and $g_{rr}^-(R)=g_{rr}^+(R)$. On the
other hand, the continuity of the second fundamental form ($K_{ij}$)
is equivalent to the O'Brien and Synge \cite{33'} junction
conditions, given as
\begin{equation*}
[G_{\gamma\delta}r^\delta]_\Sigma=0,
\end{equation*}
where $r_\gamma$ is a unit radial vector. Using the above equation
and the field equations imply $[T_{\gamma\delta}r^\delta]_\Sigma=0$
which leads to $p_r(R)=0$. Moreover, the scalar field corresponding
to the vacuum Schwarzschild solution is derived using the technique
in \cite{33''} which comes out to be $\Phi=e^{(1-\frac{2M}{r})}$. We
denote the interior and exterior regions by $\mathcal{V^-}$ and
$\mathcal{V^+}$, respectively.

The hypersurface is defined by the metric
\begin{equation}\label{50}
ds^2=d\tau^2-R^2(d\theta^2+\sin^2\theta d\phi^2),
\end{equation}
where $\tau$ is the proper time on the boundary. The extrinsic
curvature of $\Sigma$ is given by
\begin{equation}\nonumber
K_{ij}^{\pm}=-n_\gamma^{\pm}\frac{\partial^2x^\gamma_{\pm}}{\partial\eta^i\eta^j}
-n_\gamma^{\pm}\Gamma^\gamma_{\delta\mu}\frac{\partial
x^\delta_{\pm}}{\partial\eta^i} \frac{\partial
x^\mu_{\pm}}{\partial\eta^j},
\end{equation}
where $\eta^i$ are the coordinates defined on the $\Sigma$.
Moreover, the components of the four-vector normal
($n_{\gamma}^\pm$) to the hypersurface are defined in the
coordinates $(x^\gamma_{\pm})$ of $\mathcal{V}^\pm$ as
\begin{equation}\nonumber
n_{\gamma}^\pm=\pm
\frac{df}{dx^\gamma}|g^{\delta\mu}\frac{df}{dx^\delta}\frac{df}{dx^\mu}|^{\frac{-1}{2}},
\end{equation}
with $n_\gamma n^\gamma=1$. The unit normal vectors have the
following form
\begin{equation}\label{51}
n^-_\gamma=(0,e^{\frac{\lambda}{2}},0,0),\quad
n^+_\gamma=(0,(1-\frac{2M}{r})^{\frac{-1}{2}},0,0).
\end{equation}
Comparing the metrics (\ref{3}) and (\ref{49}) with (\ref{50}), it
follows that
\begin{equation}\label{56a}
[\frac{dt}{d\tau}]_\Sigma=[e^{\frac{-\nu}{2}}]_\Sigma=[(1-\frac{2M}{r})^{\frac{-1}{2}}]_\Sigma,\quad
[r]_\Sigma=R.
\end{equation}
Using Eq.(\ref{51}), the non-zero components of curvature are
calculated as
\begin{eqnarray*}
K_{00}^-&=&[-\frac{e^{-\frac{\lambda}{2}}\nu'}{2}]_\Sigma, \quad
K_{22}^-=\frac{1}{\sin^2(\theta)}K_{33}^-=[re^{-\frac{\lambda}{2}}]_\Sigma,\\
K_{00}^+&=&[-\frac{M}{r^2}(1-\frac{2M}{r})^{\frac{-1}{2}}]_\Sigma,
\quad
K_{22}^+=\frac{1}{\sin^2(\theta)}K_{33}^+=[r(1-\frac{2M}{r})^{\frac{1}{2}}]_\Sigma.
\end{eqnarray*}
The junction conditions $[K^-_{22}]_\Sigma=[K^+_{22}]_\Sigma$ and
$[r]_\Sigma=R$ yield
\begin{equation}\label{56}
e^{-\frac{\lambda(R)}{2}}=(1-\frac{2M}{R})^{\frac{1}{2}}.
\end{equation}
Substituting the above equation in the matching condition
$[K^-_{00}]_\Sigma=[K^+_{00}]_\Sigma$ gives
\begin{equation}\label{57}
\nu'(R)=\frac{2M}{R(R-2M)}.
\end{equation}
Thus, the matching conditions in Eqs.(\ref{56a})-(\ref{57}) provide
the following relations at the hypersurface
\begin{eqnarray*}
e^{\nu(R)}&=&Ce^{2Ar^2}=1-\frac{2M}{R},\\\label{15}
e^{-\lambda(R)}&=&\frac{1}{1+ADR^2e^{2AR^2}}=1-\frac{2M}{R},\\\label{16}
\nu'(R)&=&\frac{2M}{R(R-2M)}.
\end{eqnarray*}
Inserting $D=16ABC$ in the above equations, the deterministic
parameters of the system are expressed as
\begin{eqnarray}\label{17}
A&=&\frac{M}{2R^2(R-2M)},\\\label{18}
B&=&\frac{R^3}{2M},\\\label{19}
C&=&e^{\frac{M}{2M-R}}\frac{R-2M}{R},\\\label{20}
D&=&4e^{\frac{M}{2M-R}}.
\end{eqnarray}
For the metric functions in Eqs.(\ref{11}) and (\ref{11'}) along
with Eqs.(\ref{17})-(\ref{20}), the state variables are expressed in
Eqs.(\ref{21})-(\ref{23}).

\section{Physical Features of Compact Stars}

The effect of coupling parameter as well as mass of scalar field on
stellar structure can now be analyzed through the energy density and
radial/transverse pressure components. Since Gravity Probe B
experiment provides the lower bound on the mass of scalar field as
$m_\Phi>10^{-4}$ (in dimensionless units) \cite{23, 26}, we take the
values of $m_\Phi$ as $0.001$ and $0.3$. Numerical results have been
obtained for $\omega_{BD}=20,~25,~30$ which are in accordance with
the constraints imposed by the solar system observations \cite{23}.
The expression of scalar field is derived by solving the wave
equation numerically with the initial conditions
$\Phi(0)=\Phi_c=$constant and $\Phi'(0)=0$. The values of $\Phi_c$
for different values of $m_{\Phi},~\omega_{BD}$ and $\mathcal{B}$
are given in Tables \textbf{1} and \textbf{2}. All deductions have
been presented graphically for LMC X-$4$ ($M=1.29M_{\bigodot}$
\cite{33a}).

Using the condition $p_r(R)=0$, radius as well as physical
parameters of the strange star candidate are displayed in Tables
\textbf{1} and \textbf{2} for $m_\Phi=0.001$ and $m_\Phi=0.3$,
respectively. Here, the subscripts $c$ and $s$ denote that the
quantity has been calculated at the center and surface of the star,
respectively. For a physically valid solution, the metric potentials
must be positive, regular and monotonically increasing functions of
the radial coordinate \cite{34}. The potential functions are shown
in Figure \textbf{1} which reveal their regular behavior leading to
singularity free system.

The influence of physical variables such as energy density and
pressure cannot be neglected in extremely dense strange stars. The
behavior of these physical quantities with respect to the radial
coordinate is positive throughout and maximum at the center of
compact configuration as presented in Figures \textbf{2} and
\textbf{3} which shows that the core is highly concentrated for the
chosen values of the parameters
$(m_\Phi,~\omega_{BD},~\mathcal{B})$. The plots also depict the
monotonic decreasing trend of energy density and pressure components
away from the center of stars leading to a compact profile. Hence,
for the considered values of $\mathcal{B}$, the existence of quark
stars is ensured for $V(\Phi)=\frac{1}{2}m_{\Phi}^2\Phi^2$.
\begin{table}
\caption{Physical parameters of LMC X-4 with $m_\Phi=0.001$ for
different values of $\omega_{BD}$ and $\mathcal{B}$.}
\begin{center}
\begin{tabular}{|c|c|c|c|c|c|}
\hline
\multicolumn{6}{|c|}{$\mathcal{B}=64MeV/fm^3$} \\
\hline {$\omega_{BD}$} &$\Phi_c$& Predicted & $\rho_c~(gm/cm^3)$  &
$\rho_s~(gm/cm^3)$  & $p_{c}~(dyne/cm^2)$\\
&& Radius $(km)$&&& \\
\hline 20 &0.0204& $8.3141_{-0.3285}^{+0.3293}$ &
$6.6437\times10^{14}$
& $4.6155\times10^{14}$& $7.1544\times10^{34}$\\
\hline $25$ &003055&$9.6173_{-0.3847}^{+0.3863}$ &
$6.2865\times10^{14}$ &
$4.5450\times10^{14}$& $5.6021\times10^{34}$\\
\hline $30$&0.04445&$10.9515_{-0.446}^{+0.449}$ &
$6.1019\times10^{14}$ &
$4.6035\times10^{14}$ &$4.8650\times10^{34}$ \\
\hline GR limit && 5.54 & $9.498\times10^{16}$ &
$6.823\times10^{16}$ &
$3.540\times10^{36}$\\
\hline
\multicolumn{6}{|c|}{$\mathcal{B}=83MeV/fm^3$} \\
\hline {$\omega_{BD}$} &$\Phi_c$& Predicted & $\rho_c~(gm/cm^3)$  &
$\rho_s~(gm/cm^3)$  & $p_{c}~(dyne/cm^2)$\\
&& Radius $(km)$&&& \\
\hline 20 &00264& $8.3397_{-0.3315}^{+0.3582}$ &
$8.5180\times10^{14}$ &
$5.9133\times10^{14}$& $9.0542\times10^{34}$\\
\hline $25$ &0.0399&$9.6669_{-0.3908}^{+0.3931}$ &
$8.1568\times10^{14}$ &
$5.9266\times10^{14}$& $7.1989\times10^{34}$\\
\hline $30$&0.0578&$11.0422_{-0.4577}^{+0.4622}$ &
$7.9254\times10^{14}$ & $5.8838\times10^{14}$
&$6.3199\times10^{34}$\\ \hline GR limit & &5.55 &
$9.498\times10^{16}$ & $6.823\times10^{16}$ & $3.564\times10^{36}$\\
\hline
\end{tabular}
\end{center}
\end{table}
\begin{table} \caption{Physical parameters of LMC X-4 with $m_\Phi=0.3$ for different
values of $\omega_{BD}$ and $\mathcal{B}$.}
\begin{center}
\begin{tabular}{|c|c|c|c|c|c|}
\hline
\multicolumn{6}{|c|}{$\mathcal{B}=64MeV/fm^3$} \\
\hline {$\omega_{BD}$} &$\Phi_c$& Predicted & $\rho_c~(gm/cm^3)$  &
$\rho_s~(gm/cm^3)$  & $p_{c}~(dyne/cm^2)$\\
&& Radius $(km)$&&& \\
\hline 20 & 0.0204 & $8.3141_{-0.3285}^{+0.3293}$ &
$6.7762\times10^{14}$
& $4.7467\times10^{14}$& $7.1905\times10^{34}$\\
\hline $25$ & 0.03055 & $9.6173_{-0.3847}^{+0.3863}$ &
$6.6665\times10^{14}$ &
$4.9018\times10^{14}$& $5.8065\times10^{34}$\\
\hline $30$& 0.04555 & $10.9515_{-0.446}^{+0.449}$ &
$6.7427\times10^{14}$ &
$5.1681\times10^{14}$ &$4.9215\times10^{34}$ \\
\hline
\multicolumn{6}{|c|}{$\mathcal{B}=83MeV/fm^3$} \\
\hline {$\omega_{BD}$} & $\Phi_c$ & Predicted & $\rho_c~(gm/cm^3)$
&
$\rho_s~(gm/cm^3)$  & $p_{c}~(dyne/cm^2)$\\
& &Radius $(km)$&&& \\
\hline 20 & 0.0264 & $8.3397_{-0.3315}^{+0.3582}$ &
$8.7829\times10^{14}$
& $6.1661\times10^{14}$& $9.1757\times10^{34}$\\
\hline $25$ & 0.0399 & $9.6669_{-0.3908}^{+0.3931}$ &
$8.5916\times10^{14}$ &
$6.3427\times10^{14}$& $7.2289\times10^{34}$\\
\hline $30$& 0.0578& $11.0422_{-0.4577}^{+0.4622}$ &
$8.8271\times10^{14}$ & $6.9153\times10^{14}$
&$6.0806\times10^{34}$\\ \hline
\end{tabular}
\end{center}
\end{table}
\begin{figure}\center
\epsfig{file=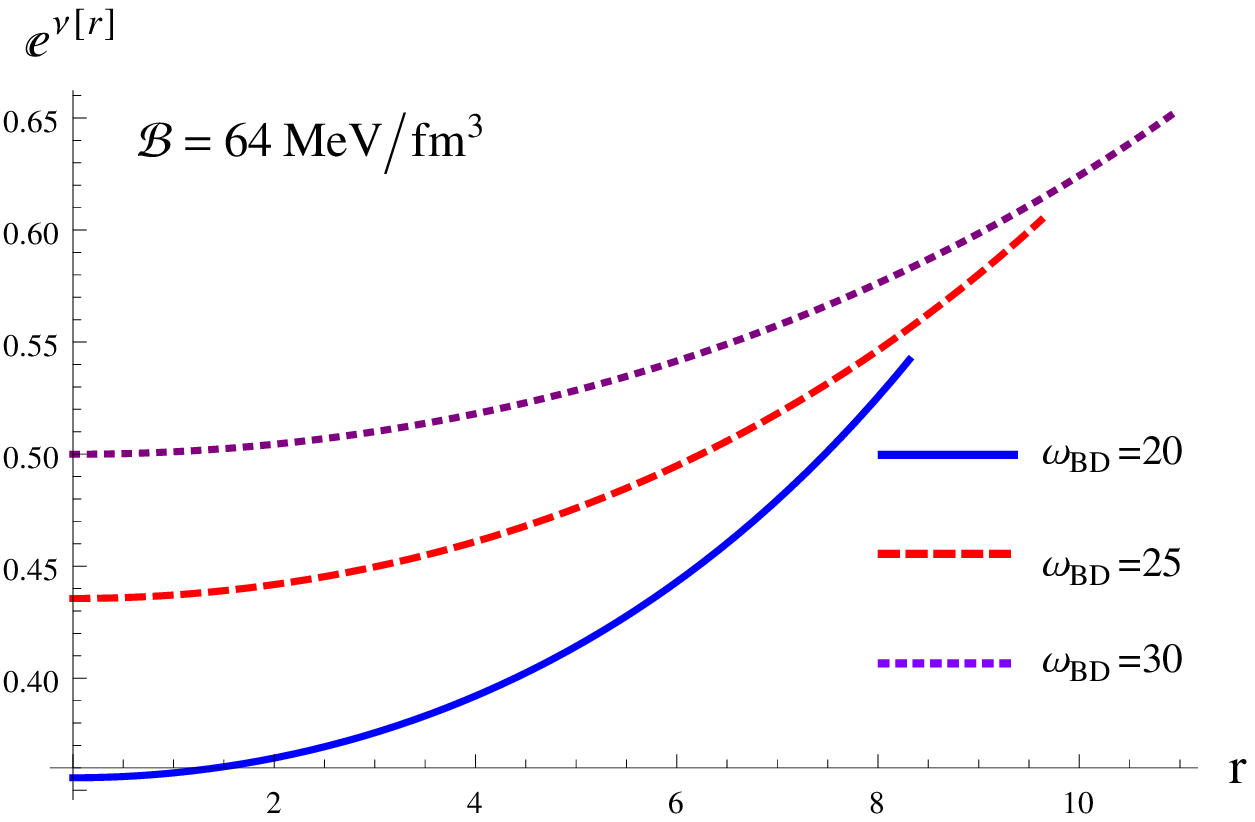,width=0.4\linewidth}\epsfig{file=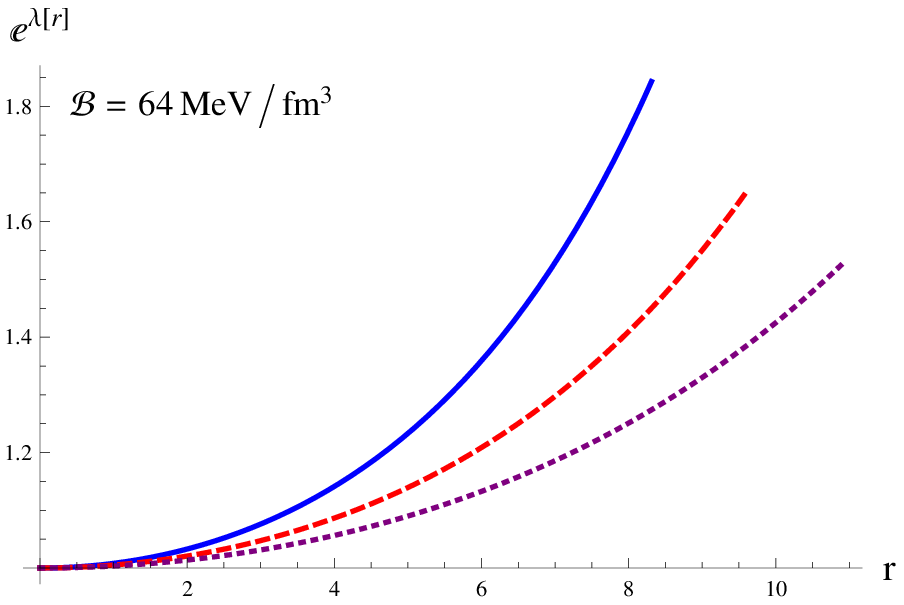,width=0.4\linewidth}
\epsfig{file=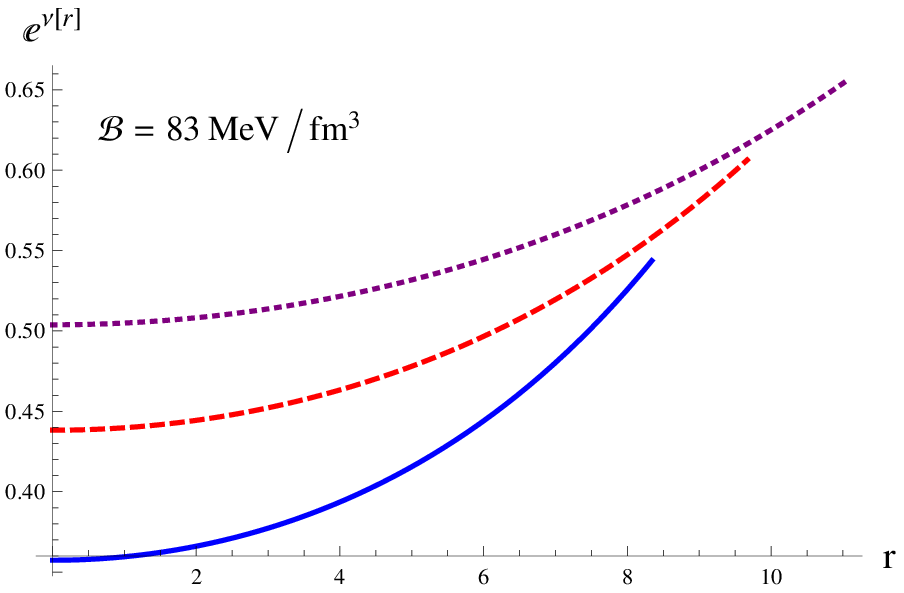,width=0.4\linewidth}\epsfig{file=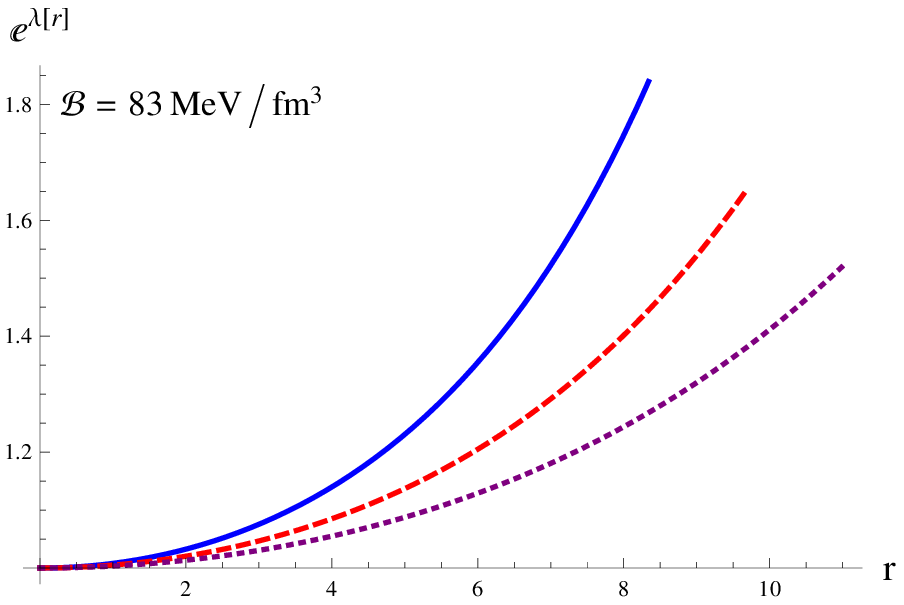,width=0.4\linewidth}\\
\caption{Plots of metric potentials for massive scalar field versus
radial coordinate.}
\end{figure}
\begin{figure}\center
\epsfig{file=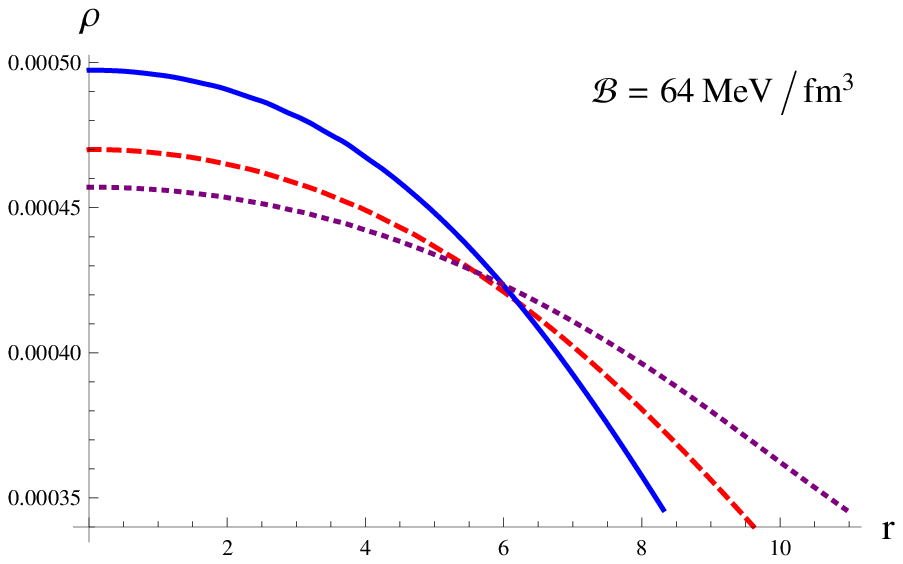,width=0.35\linewidth}\epsfig{file=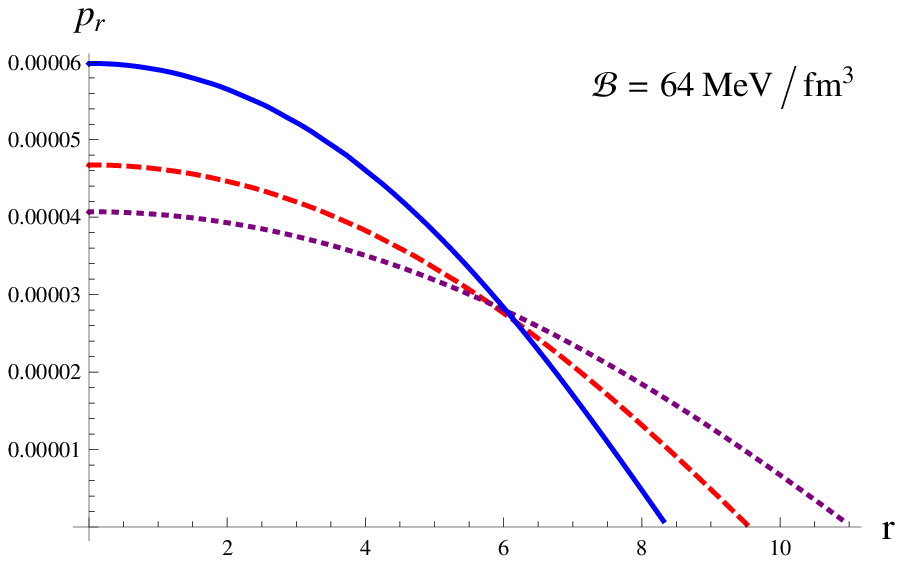,width=0.35\linewidth}\epsfig{file=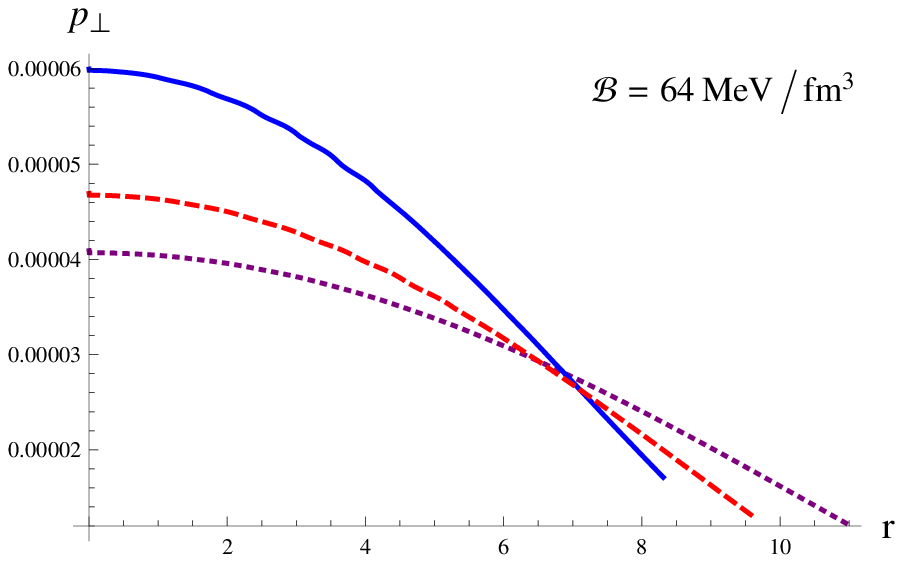,width=0.35\linewidth}
\epsfig{file=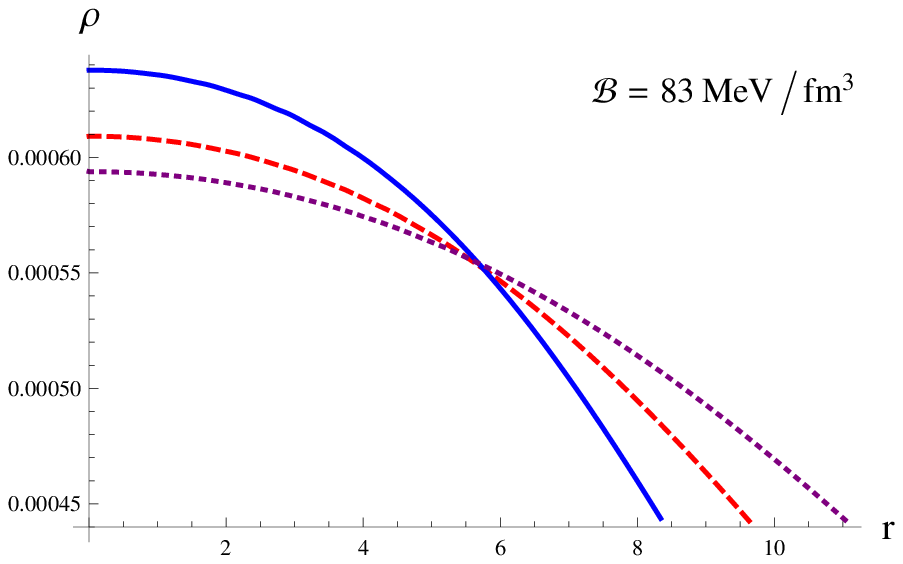,width=0.35\linewidth}\epsfig{file=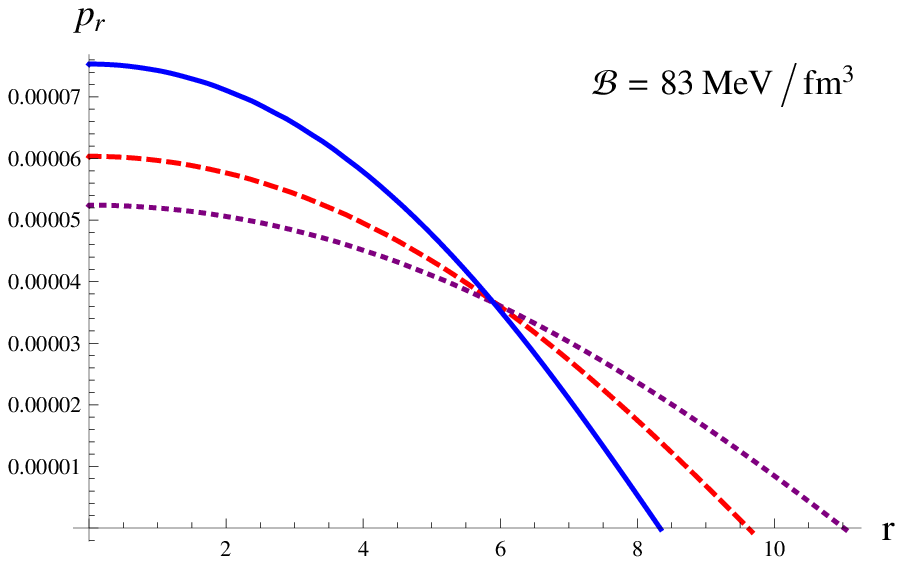,width=0.35\linewidth}\epsfig{file=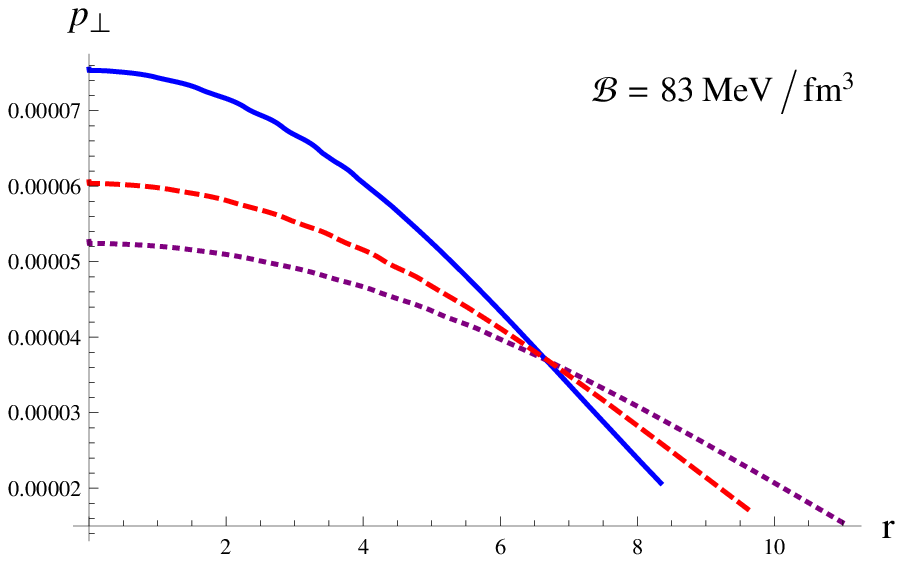,width=0.35\linewidth}
\caption{Effective energy density, effective radial/transverse
pressure as functions of $r$ with $m_\Phi=0.001$. }
\end{figure}
\begin{figure}\center
\epsfig{file=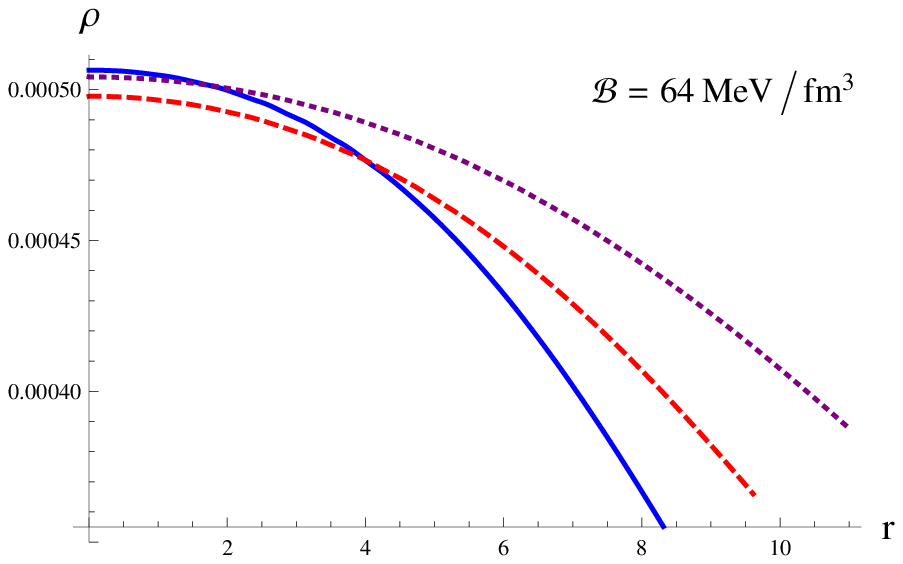,width=0.35\linewidth}\epsfig{file=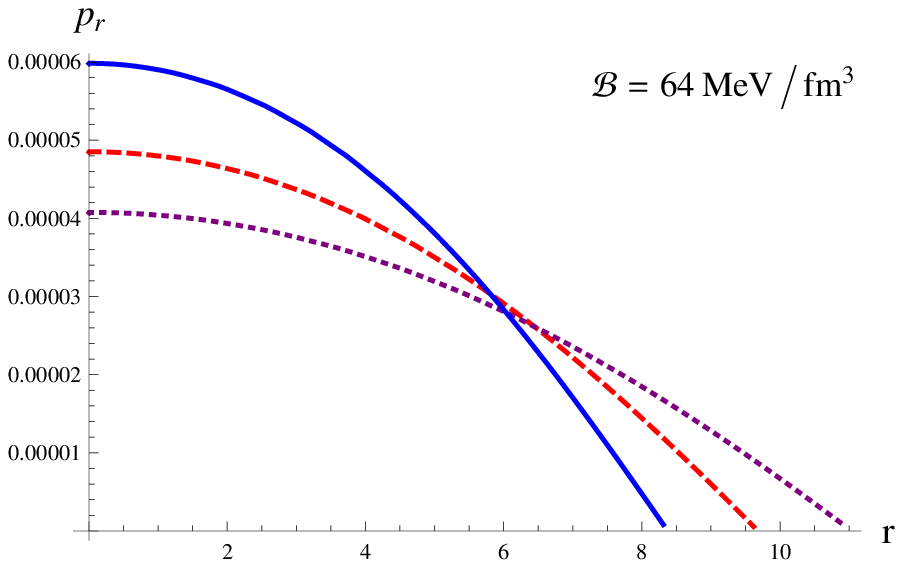,width=0.35\linewidth}\epsfig{file=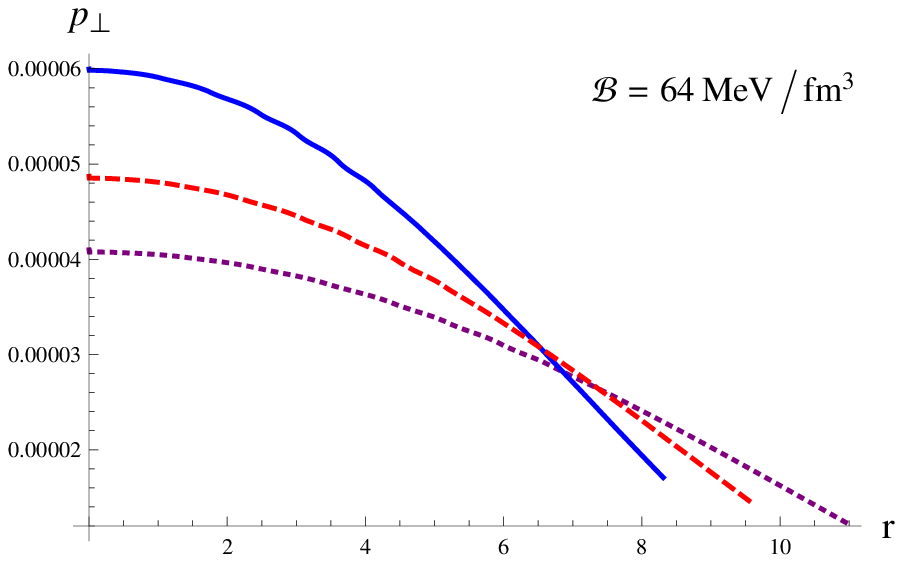,width=0.35\linewidth}
\epsfig{file=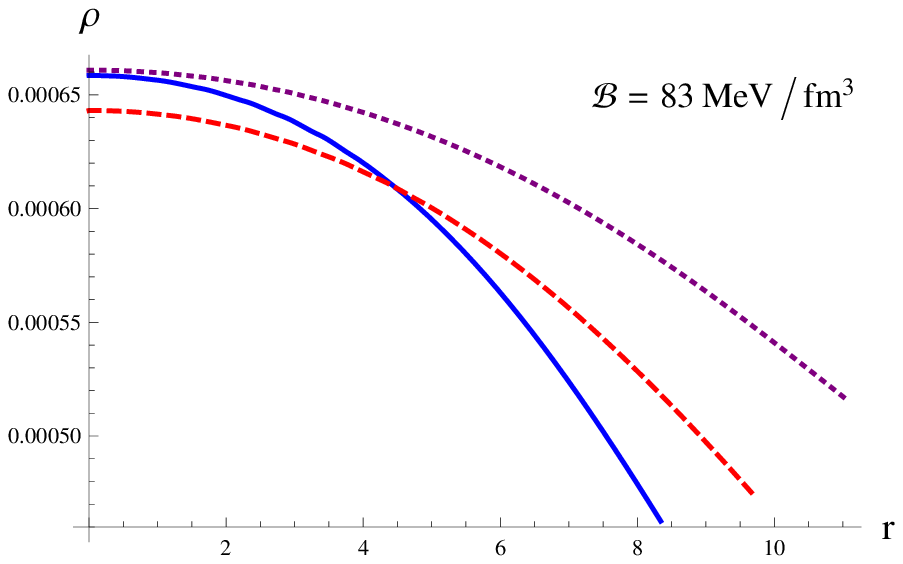,width=0.35\linewidth}\epsfig{file=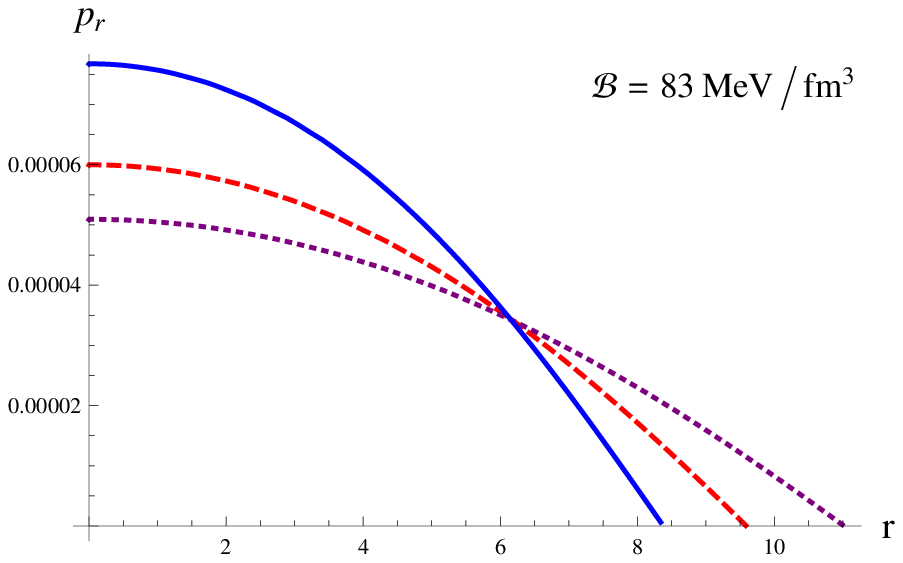,width=0.35\linewidth}\epsfig{file=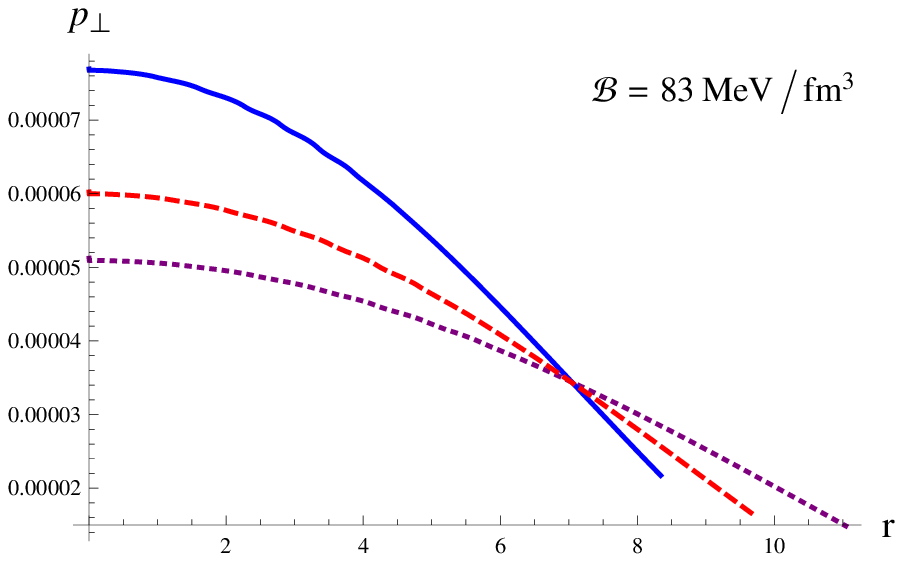,width=0.35\linewidth}
\caption{Effective energy density, effective radial/transverse
pressure as functions of $r$ with $m_\Phi=0.3$.}
\end{figure}\begin{figure}\center
\epsfig{file=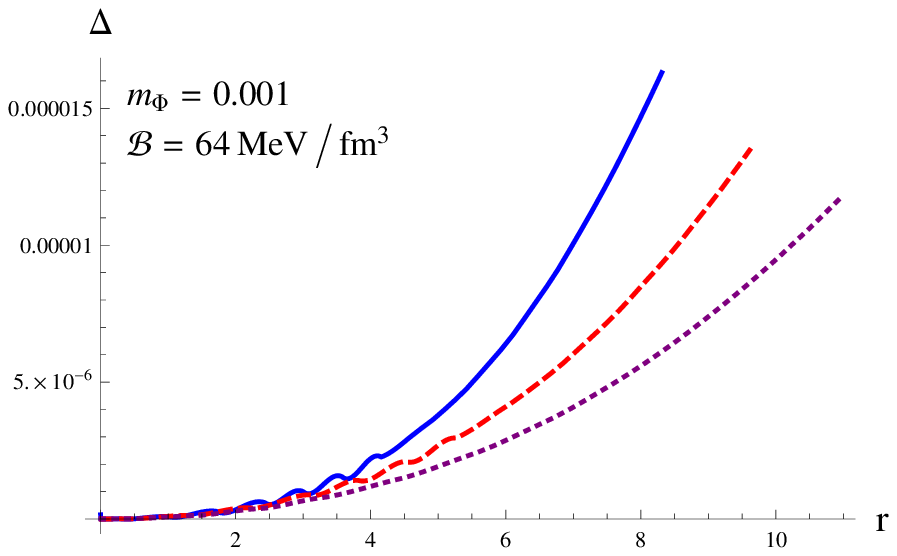,width=0.4\linewidth}\epsfig{file=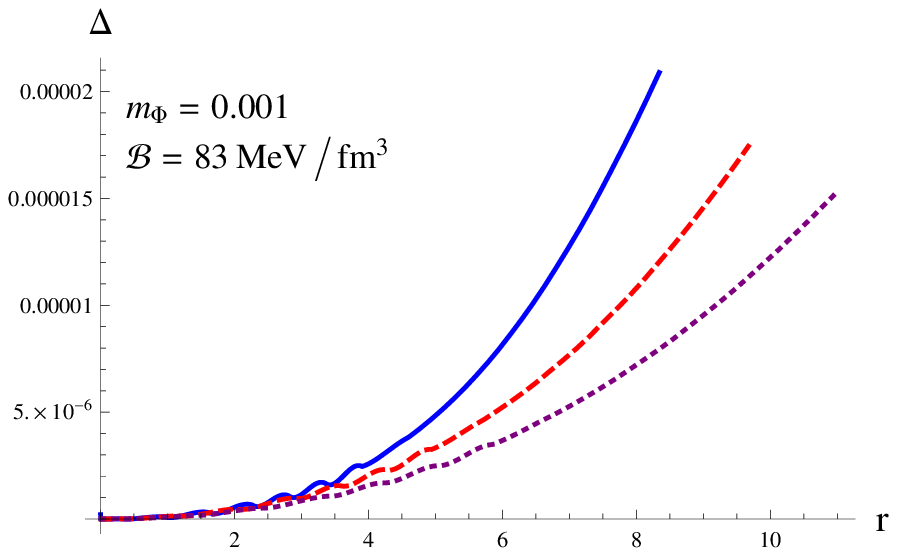,width=0.4\linewidth}\\
\epsfig{file=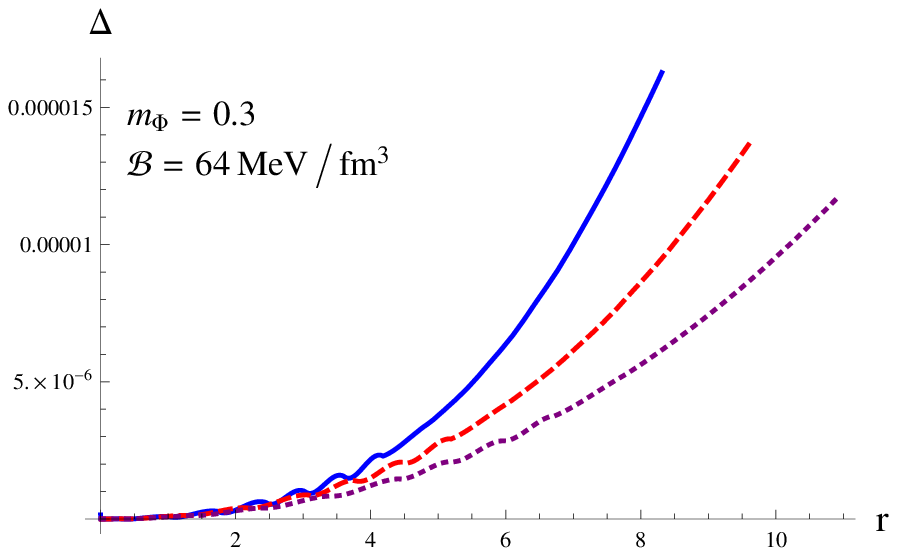,width=0.4\linewidth}\epsfig{file=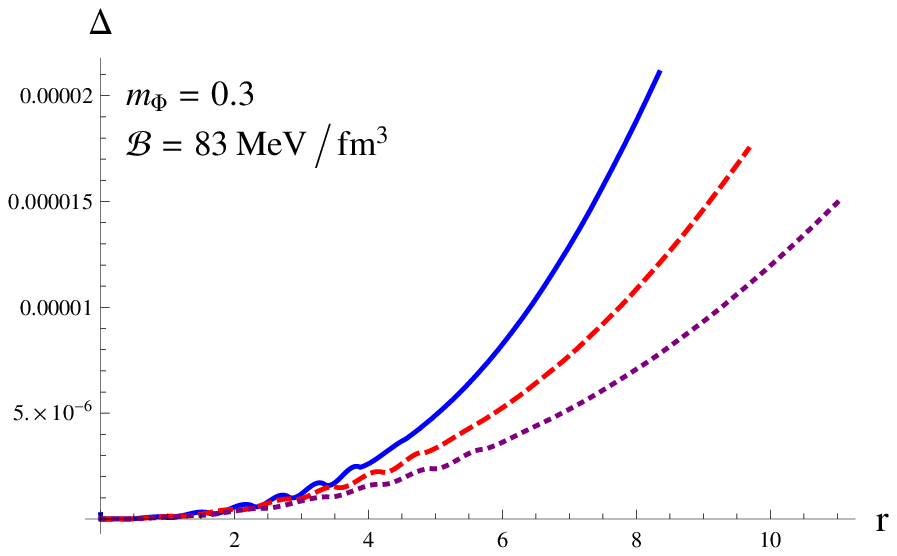,width=0.4\linewidth}
\caption{Variation of effective anisotropy as a function of $r$.}
\end{figure}

The radial and tangential components of pressure give rise to
anisotropy within the structure. The anisotropy of pressure,
measured as $\Delta=p_\perp-p_r$, is positive when $p_\perp>p_r$ and
negative otherwise. The positive and increasing behavior of
anisotropy suggests that an outward directed repelling force is in
play in the interior of stellar models stabilizing the system
against gravity. Utilizing Eqs.(\ref{22}) and (\ref{23}), the
anisotropy comes out to be
\begin{eqnarray*}
\Delta&=&\frac{\xi\Phi}{r}\left[M^2r^3\Phi^2(r)\left(1-2e^{\frac{M
(r-R)(r+R)}{R^2(R-2 M)}}\right)^2-r
R^2\omega_{BD}(2M-R)\right.\\
&\times&\left.\left(2Mr^2e^{\frac{M(r-R)(r+R)}{R^2(R-2M)}}+R^2
(R-2M)\right)\Phi'^2(r)+\Phi(r)\left(R^4\left(-(R-2
M)^2\right)\right.\right.\\
&\times&\left.\left.\left(\Phi'(r)-r\Phi''(r)\right)-2
Mr^2e^{\frac{M(r-R)(r+R)}{R^2
(R-2M)}}\left(\left(Mr^2-4MR^2+2R^3\right)\Phi'(r)\right.\right.\right.\\
&+&\left.\left.\left.rR^2(2M-R)\Phi''(r)\right)\right)\right].
\end{eqnarray*}
Figure \textbf{4} indicates that the behavior of anisotropy is
acceptable for the selected model.
\begin{figure}\center
\epsfig{file=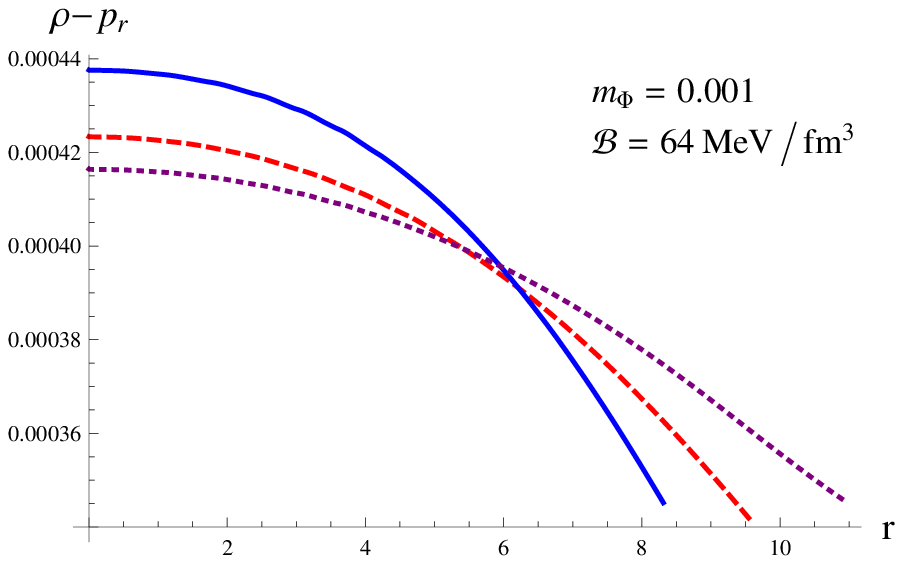,width=0.45\linewidth}\epsfig{file=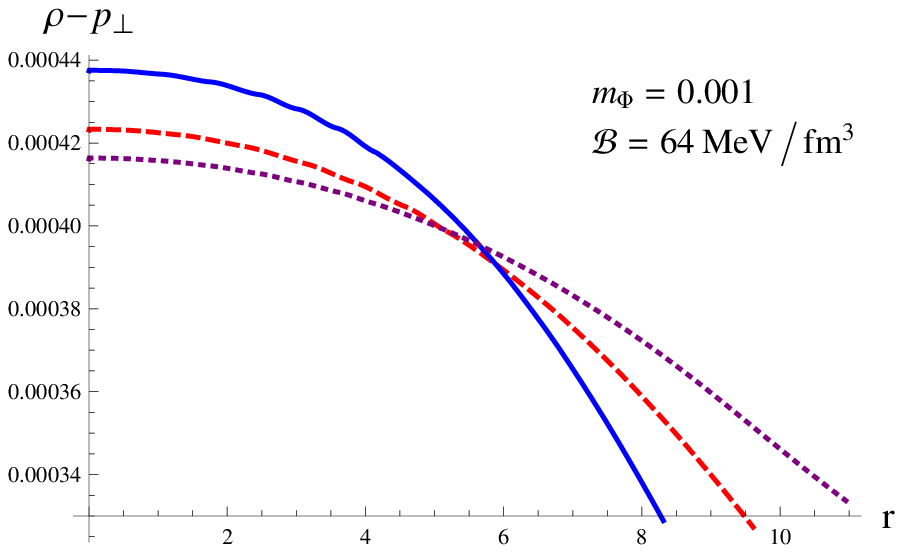,width=0.45\linewidth}\\
\epsfig{file=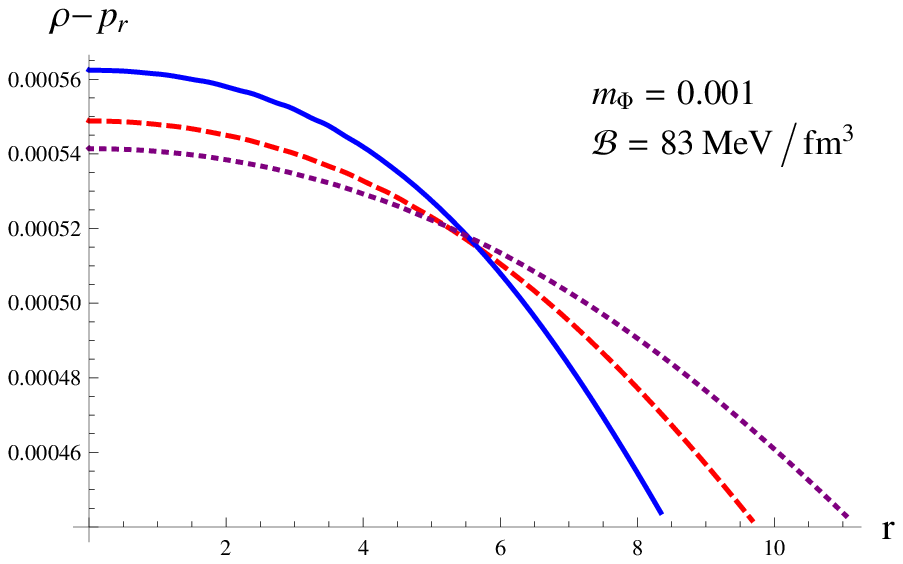,width=0.45\linewidth}\epsfig{file=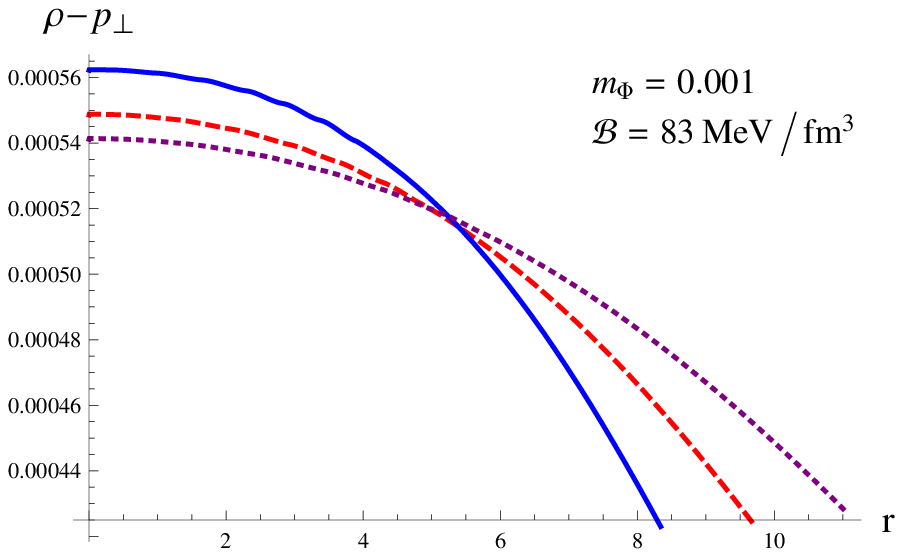,width=0.45\linewidth}\\
\epsfig{file=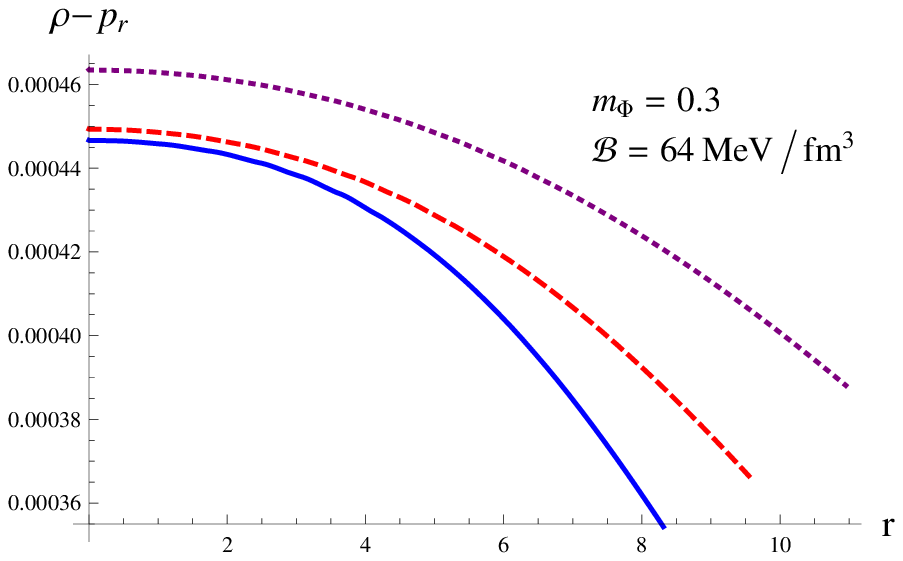,width=0.45\linewidth}\epsfig{file=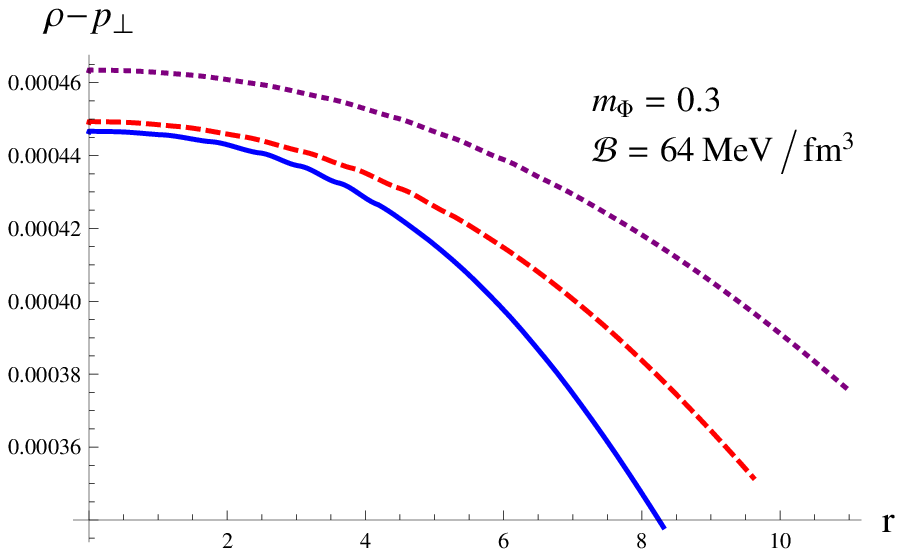,width=0.45\linewidth}\\
\epsfig{file=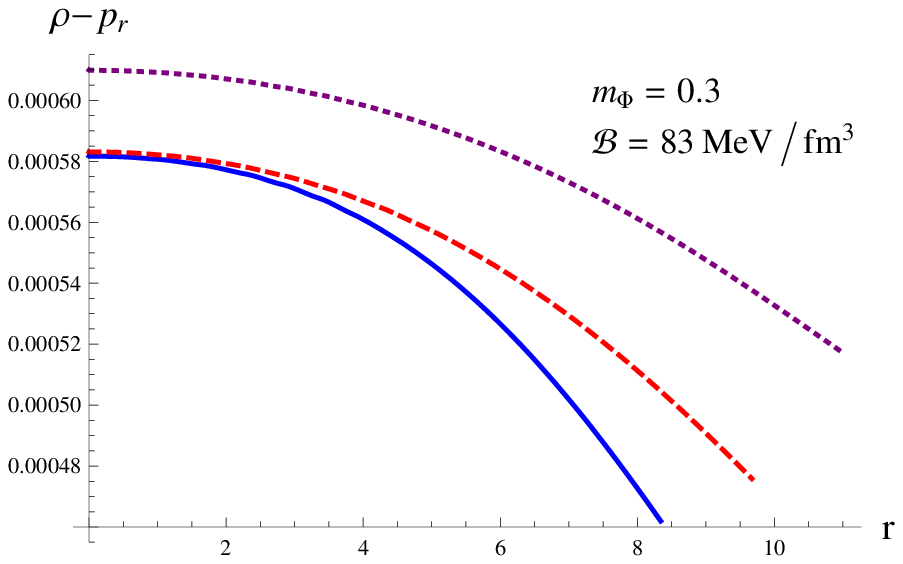,width=0.45\linewidth}\epsfig{file=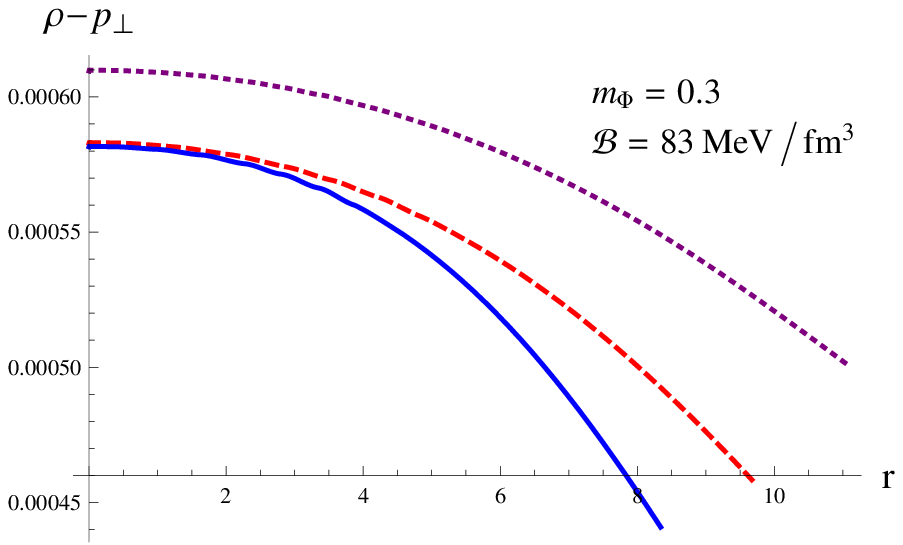,width=0.45\linewidth}
\caption{Dominant energy condition plotted against the radial
coordinate.}
\end{figure}
\begin{figure}\center
\epsfig{file=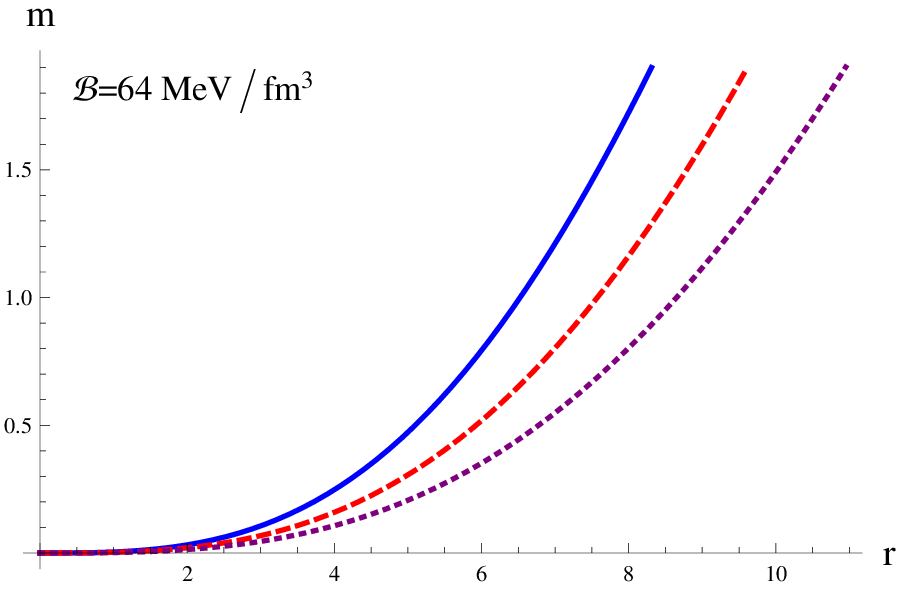,width=0.32\linewidth}\epsfig{file=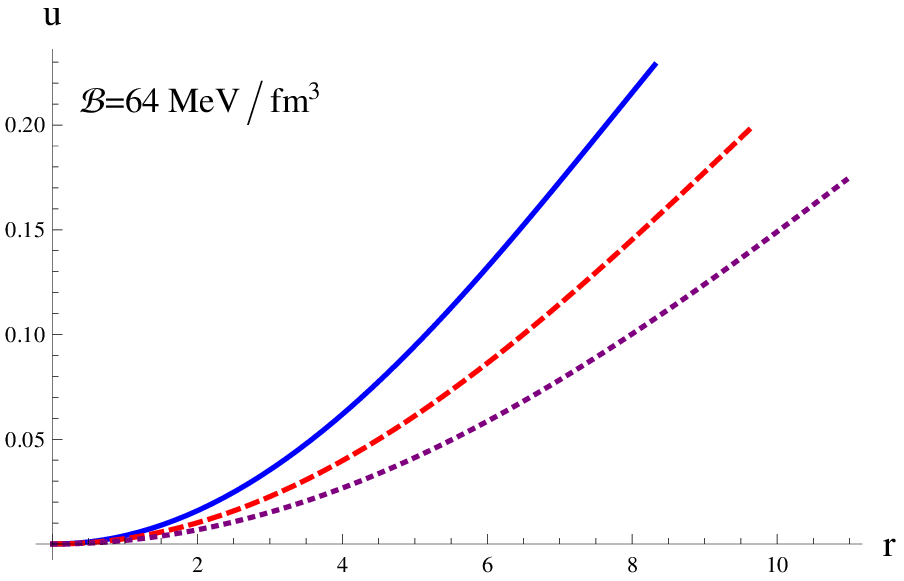,width=0.32\linewidth}\epsfig{file=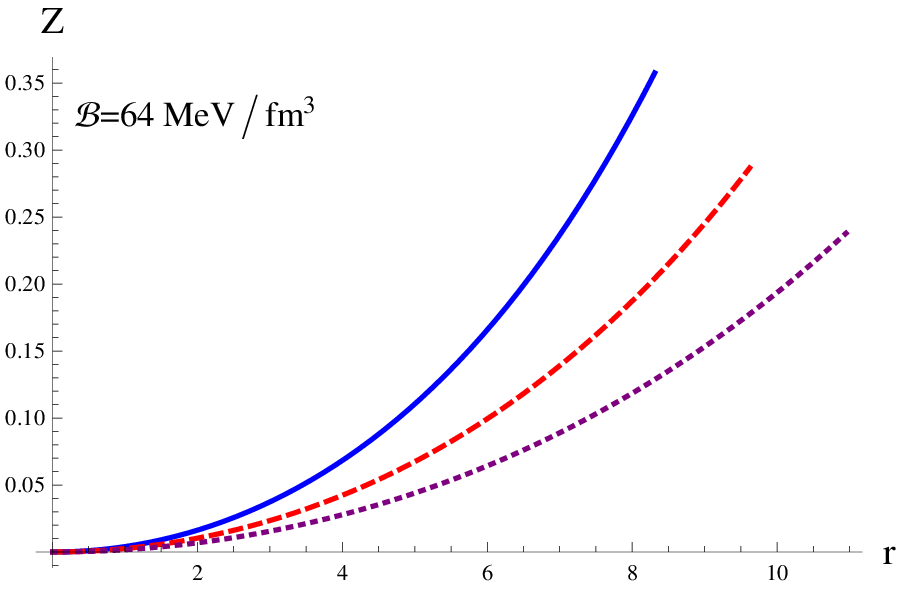,width=0.32\linewidth}\\
\epsfig{file=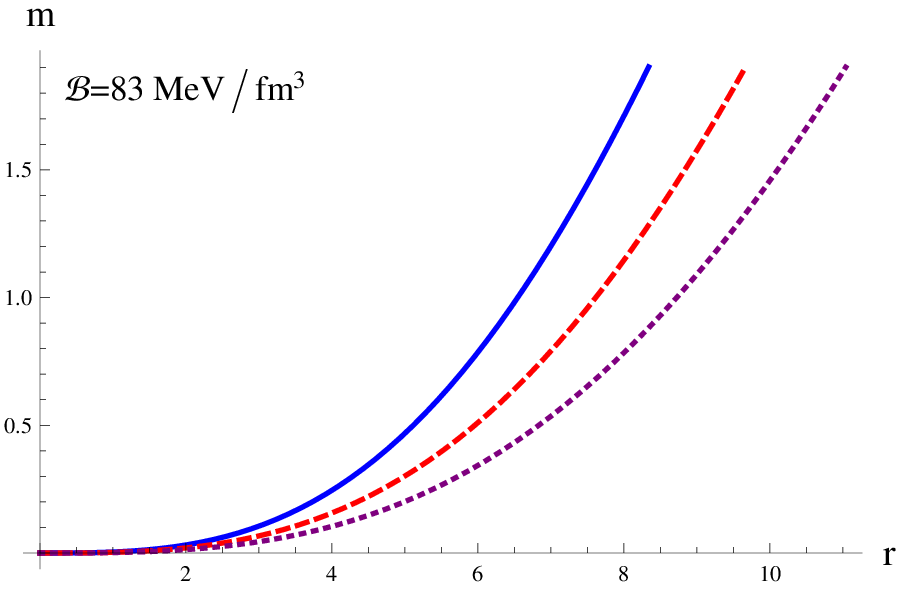,width=0.32\linewidth}\epsfig{file=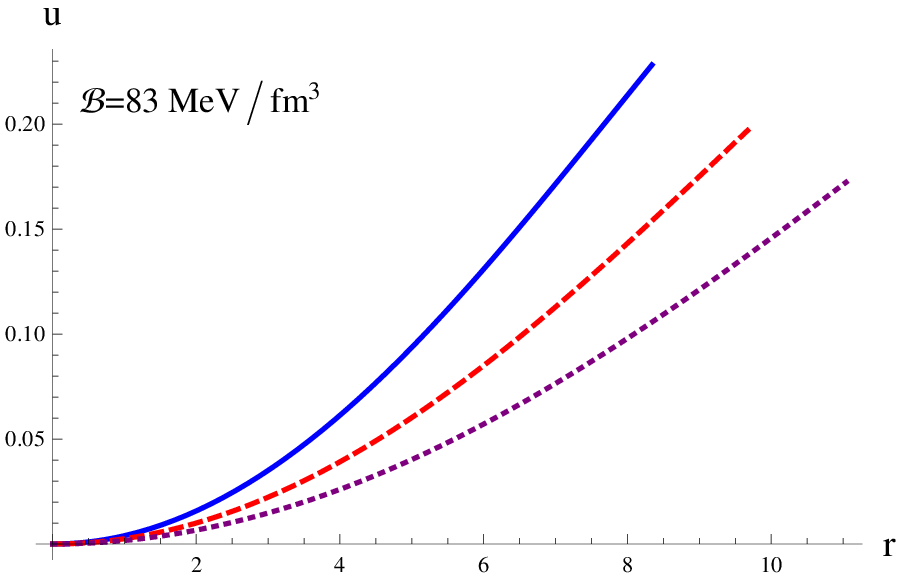,width=0.32\linewidth}\epsfig{file=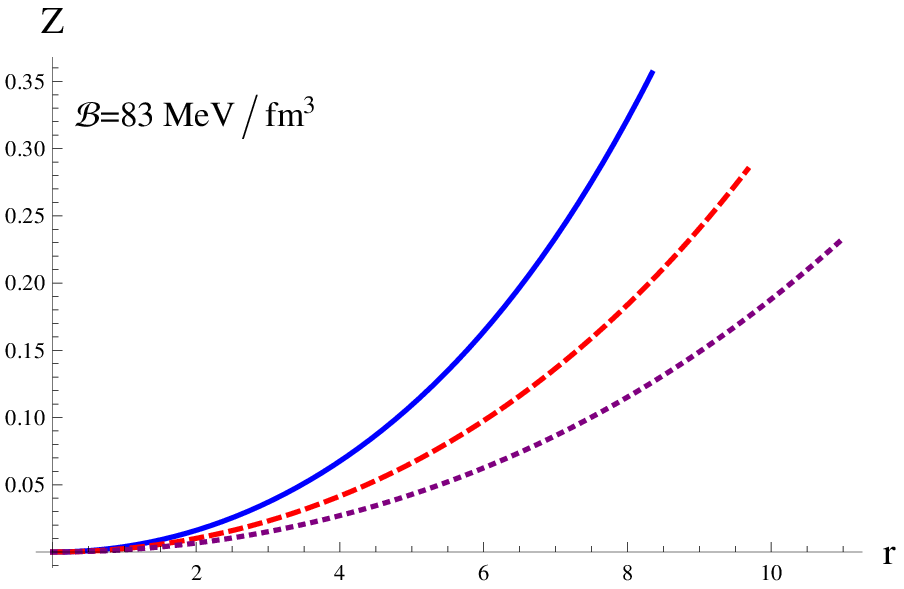,width=0.32\linewidth}\\
\caption{Plots of relation between mass, compactness factor and
redshift against radial coordinate.}
\end{figure}

\subsection{Energy Conditions}

A configuration is said to be realistic if it satisfies all four
energy conditions, i.e., null (NEC), weak (WEC), strong (SEC) and
dominant (DEC). These conditions are evaluated in terms of effective
energy density and effective components of pressure in the presence
of scalar field as \cite{35}
\begin{eqnarray*}
&&\text{NEC:}\quad\rho\geq0,\\
&&\text{WEC:}\quad\rho+p_r\geq0,\quad\rho+p_\perp\geq0,\\
&&\text{SEC:}\quad\rho+p_r+2p_\perp\geq0,\\
&&\text{DEC:}\quad\rho-p_r\geq0,\quad \rho-p_\perp\geq0.
\end{eqnarray*}
Figures \textbf{2} and \textbf{3} depict positive behavior of
$\rho,~p_r$ and $p_\perp$ throughout the stellar structure, the
first three conditions are readily satisfied. The plot of DEC in
Figure \textbf{5} is positive at each point within the stellar
structure. Hence, all energy conditions are satisfied which validate
the model for the chosen values of $m_\Phi,~\mathcal{B}$ and
$\omega_{BD}$.

\subsection{Effective Mass, Compactness and Redshift}

The size and mass are two inter-related observable features of a
compact object. The effective mass for the current structure is
calculated via Eq.(\ref{9'}) as
\begin{equation*}
m(r)=\frac{r}{2}\left[\frac{2Mr^2e^{\frac{M(R^2-r^2)}{R^2(2M-R)}}}{R^2(R-2M)
+2Mr^2e^{\frac{M(R^2-r^2)}{R^2(2M-R)}}}\right],
\end{equation*}
which is dependent on the radius of celestial body. Figure
\textbf{6} shows a decrease in mass for a larger value of
$\mathcal{B}$. The compactness function is the ratio of mass to
radius given as
\begin{equation*}
u(r)=\frac{m(r)}{r}=\frac{1}{2}\left[\frac{2Mr^2e^{\frac{M(R^2-r^2)}{R^2(2M-R)}}}{R^2(R-2M)
+2Mr^2e^{\frac{M(R^2-r^2)}{R^2(2M-R)}}}\right].
\end{equation*}
Figure \textbf{6} displays the compactness factor as a monotonic
increasing function with respect to the radial coordinate. The
values attained by the function adhere to the upper limit
$\frac{m}{R}<\frac{4}{9}$, proposed by Buchdal \cite{36} for both
values of bag constant. Further, the gravitational redshift is a
measure of the force exerted on light as a consequence of strong
gravity. The relativistic effect can be measured from the X-ray
spectrum of the cosmic object using the compactness factor which is
defined as
\begin{equation*}
Z=\frac{1}{\sqrt{1-2u(r)}}-1,
\end{equation*}
leading to the following expression
\begin{equation*}
Z=-1+\sqrt{1+\frac{2Mr^2e^{\frac{M(R^2-r^2)}{R^2(2M-R)}}}{R^2(R-2M)}}.
\end{equation*}
Figure \textbf{6} exhibits the redshift as an increasing function of
radial coordinate. We would like to mention here that the surface
redshift for the stellar candidate is consistent with the limit for
relativistic stars ($Z<5.211$) \cite{37}.

\subsection{Stability of Stellar System}
\begin{figure}\center
\epsfig{file=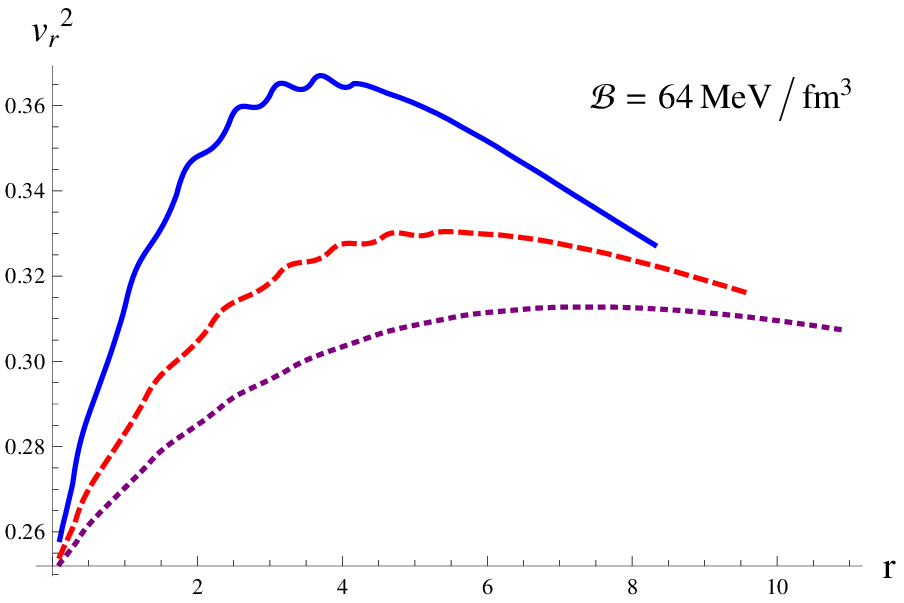,width=0.32\linewidth}\epsfig{file=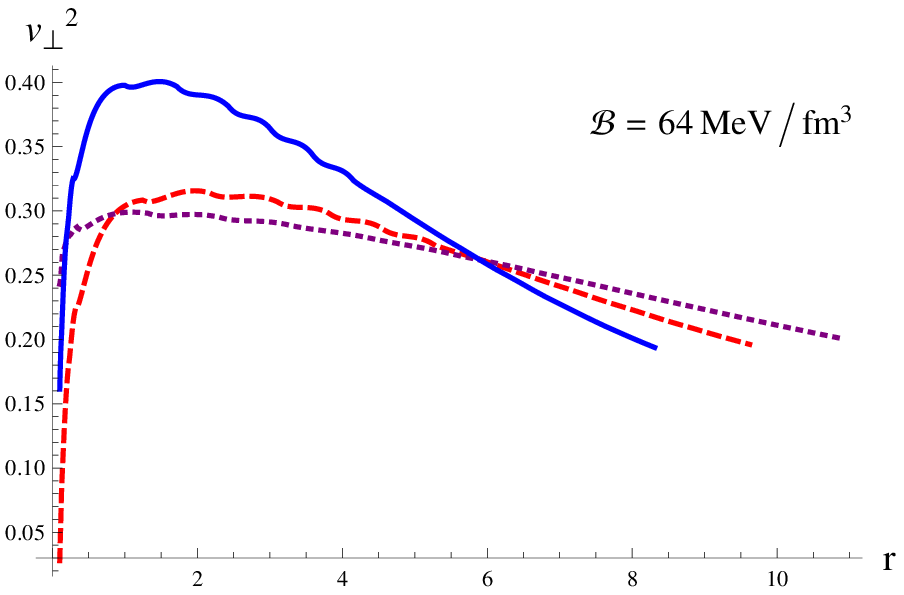,width=0.32\linewidth}\epsfig{file=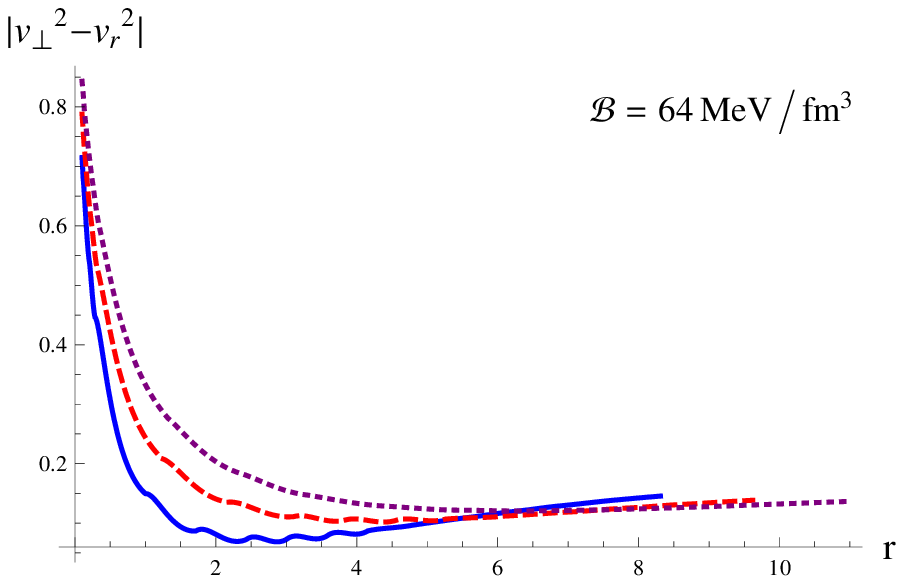,width=0.32\linewidth}\\
\epsfig{file=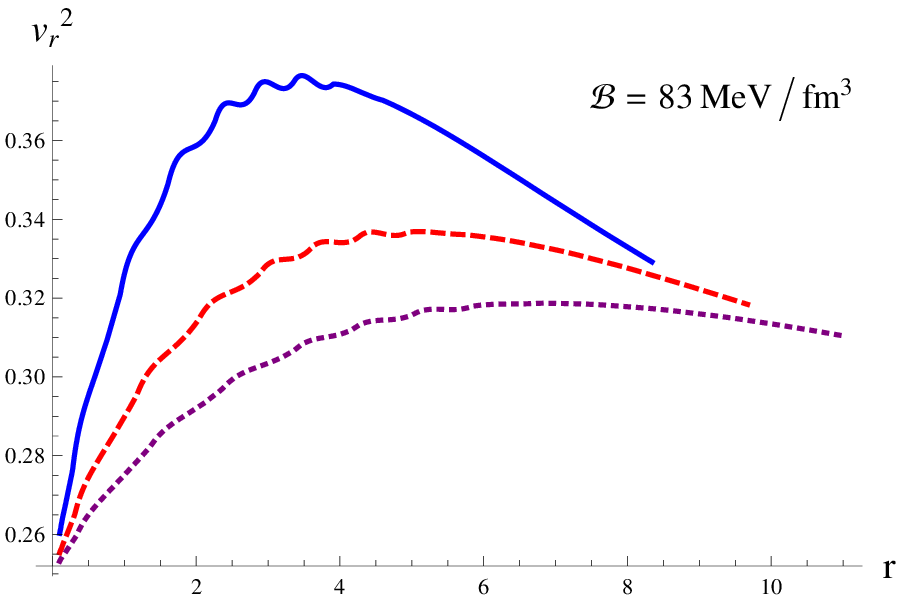,width=0.32\linewidth}\epsfig{file=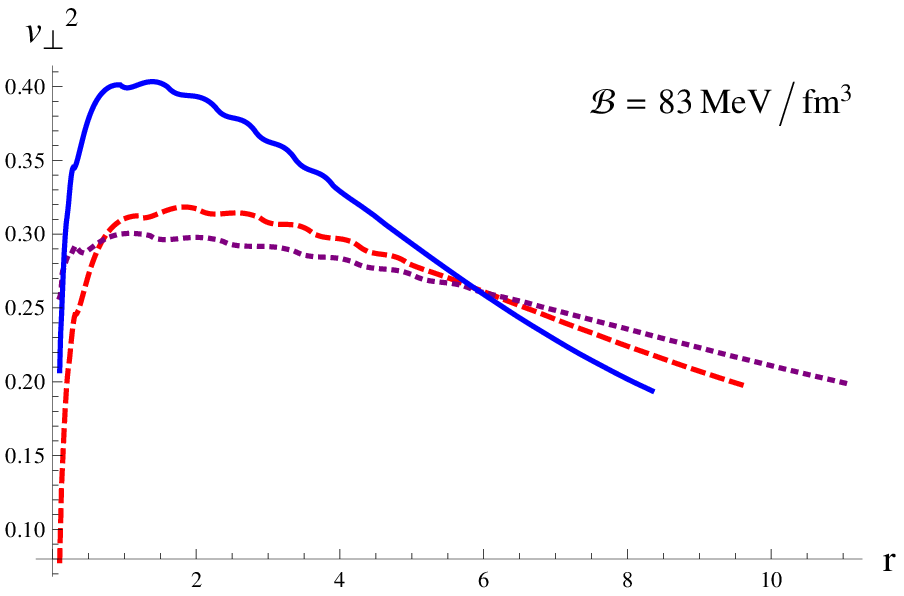,width=0.32\linewidth}\epsfig{file=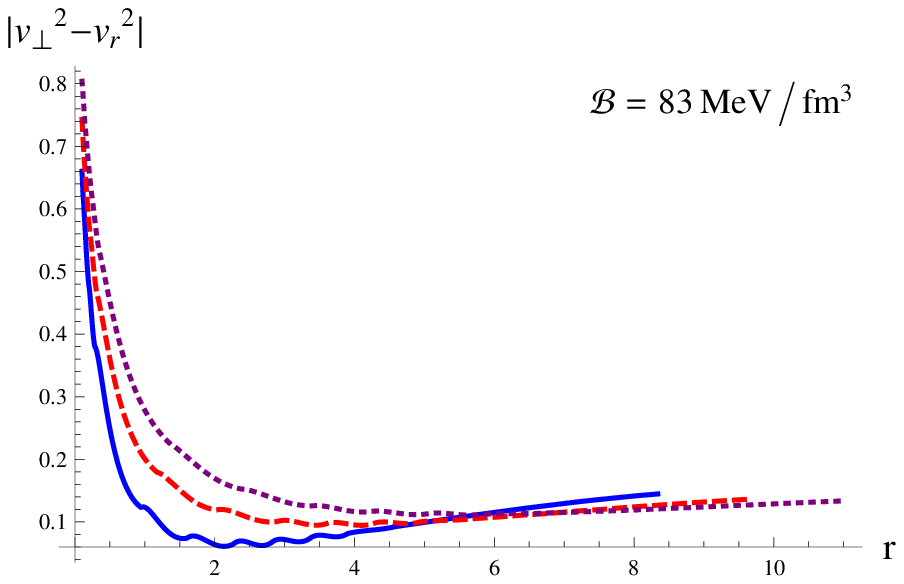,width=0.32\linewidth}\\
\caption{Variation of radial velocity, tangential velocity and
$|v_\perp^2-v_r^2|$ with respect to radial coordinate with
$m_\Phi=0.001$.}
\end{figure}
In this section, we examine stability of the anisotropic setup. It
is crucial for a stable anisotropic system that the speed of sound
is less than that of light, i.e., $0<v_r^2<1$ and $0<v_\perp^2<1$,
where $v_r$ and $v_\perp$ are the radial and tangential components
of speed expressed, respectively, as
\begin{equation*}
v_r^2=\frac{dp_r}{d\rho},\quad v_\perp^2=\frac{dp_\perp}{d\rho}.
\end{equation*}
This criterion is known as the condition of causality \cite{38}. The
stability of a system can also be verified through Herrera's
cracking approach \cite{39}. Cracking occurs when inward directed
radial forces of a perturbed system change direction for some value
of radial coordinate. According to this scheme, a region free from
cracking is stable when $0<|v_\perp^2-v_r^2|<1$. One of the
interesting features of this method is that cracking is closely
related to changes in local anisotropy. Figures \textbf{7} and
\textbf{8} shows that anisotropic distribution agrees with the
causality condition as well as cracking approach in the framework of
MBD theory.
\begin{figure}\center
\epsfig{file=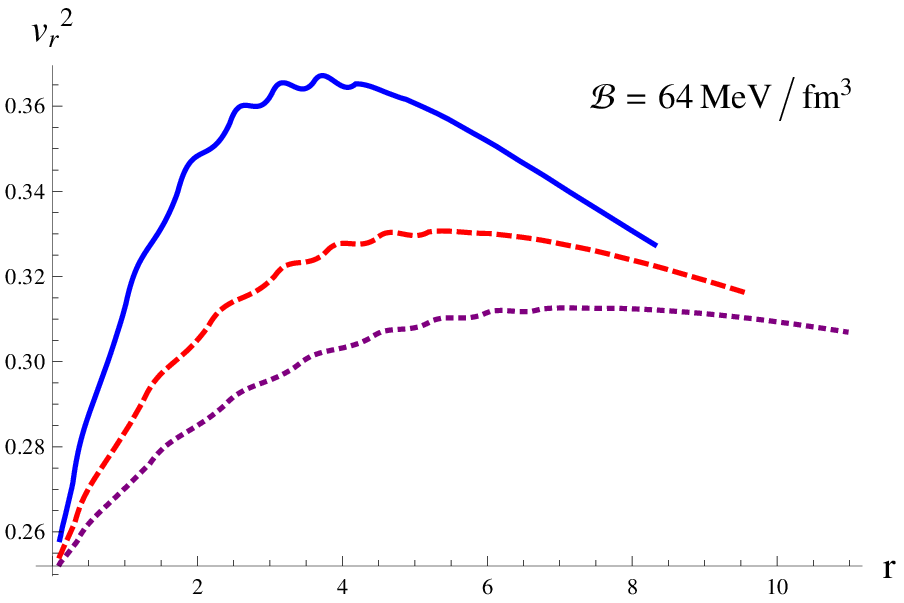,width=0.32\linewidth}\epsfig{file=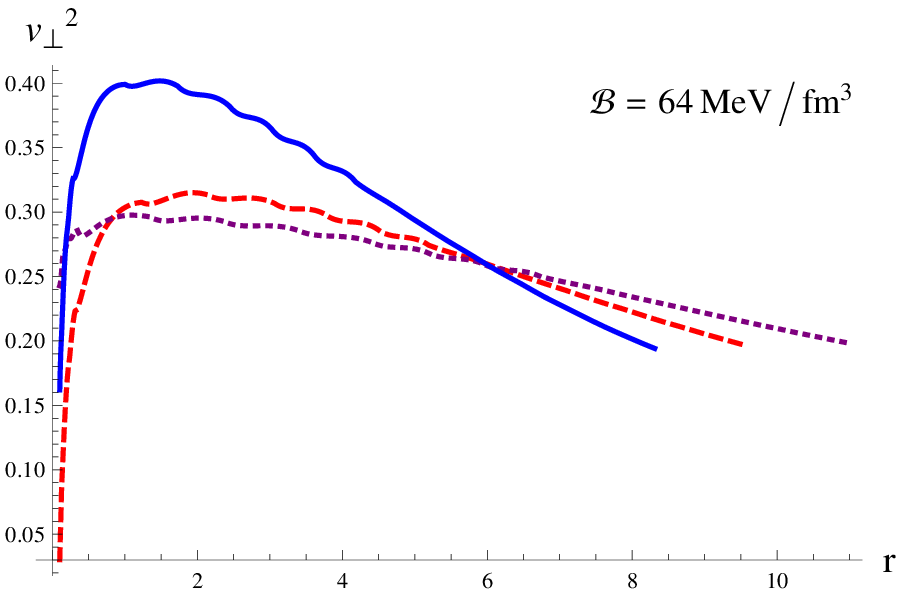,width=0.32\linewidth}\epsfig{file=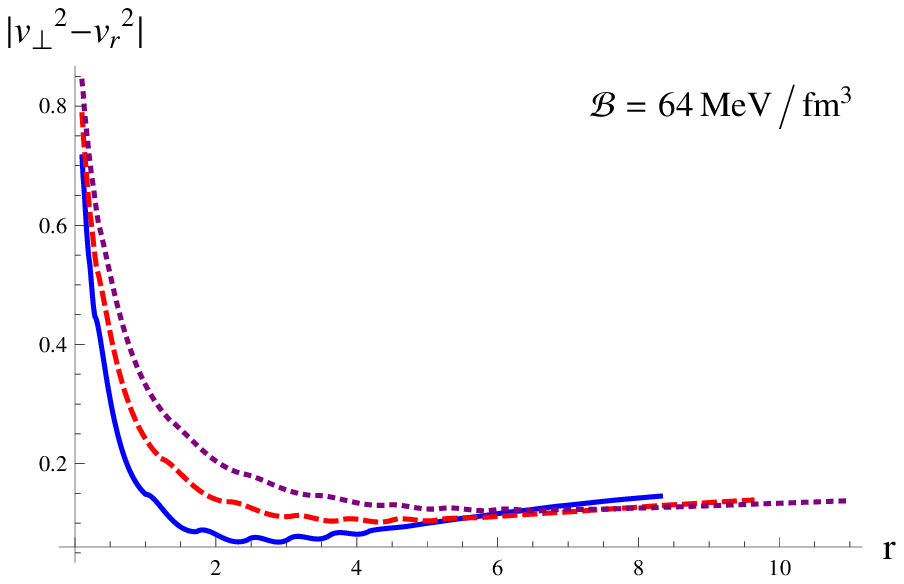,width=0.32\linewidth}\\
\epsfig{file=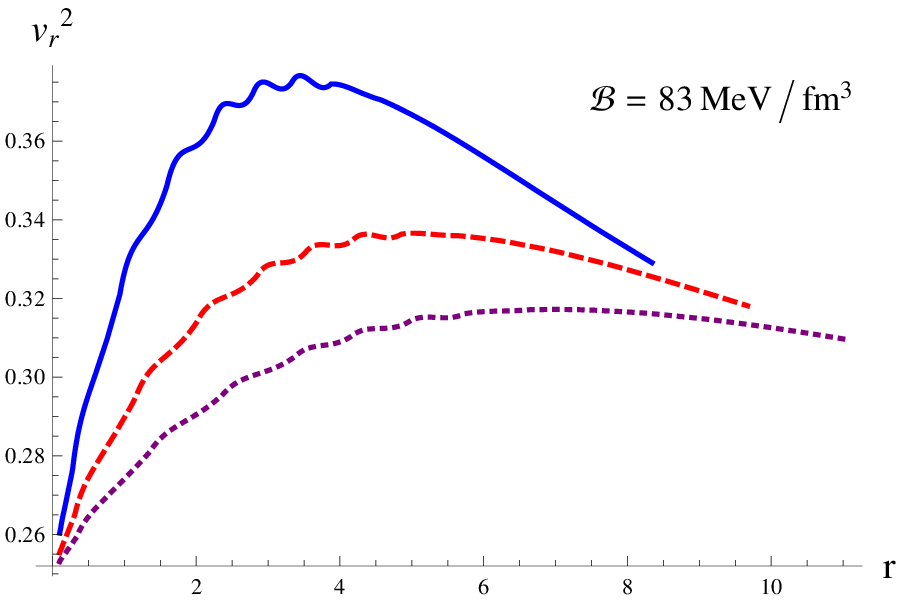,width=0.32\linewidth}\epsfig{file=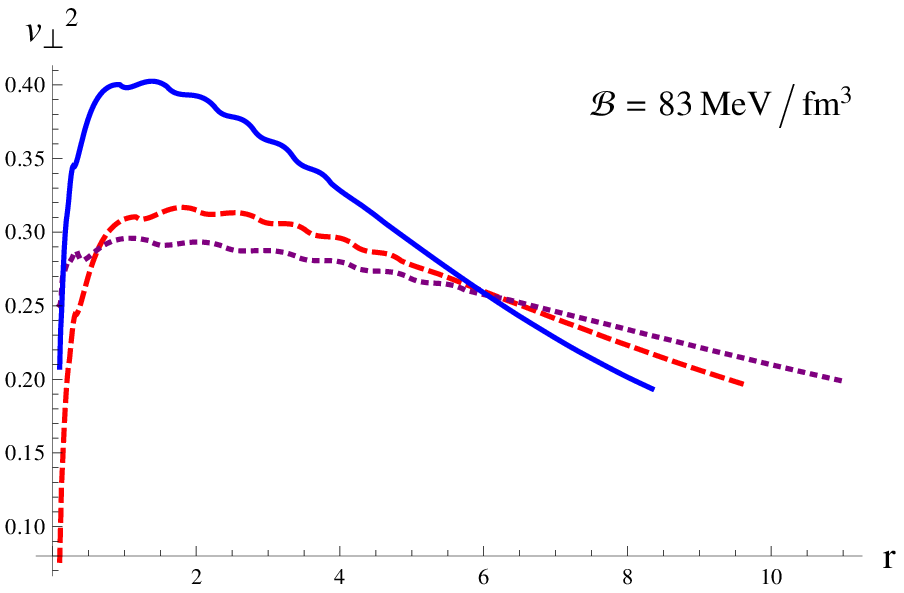,width=0.32\linewidth}\epsfig{file=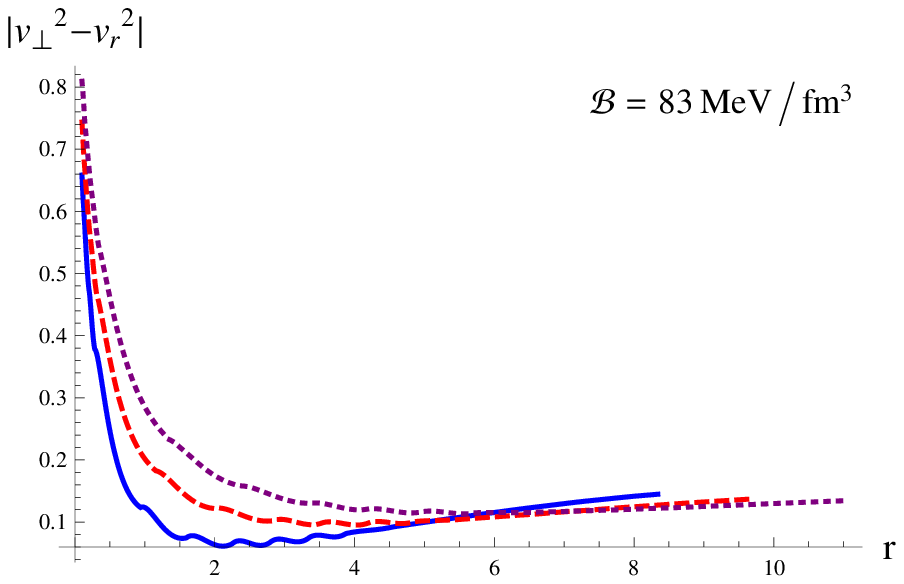,width=0.32\linewidth}
\caption{Variation of radial velocity, tangential velocity and
$|v_\perp^2-v_r^2|$ with respect to radial coordinate with
$m_\Phi=0.3$.}
\end{figure}
\begin{figure}\center
\epsfig{file=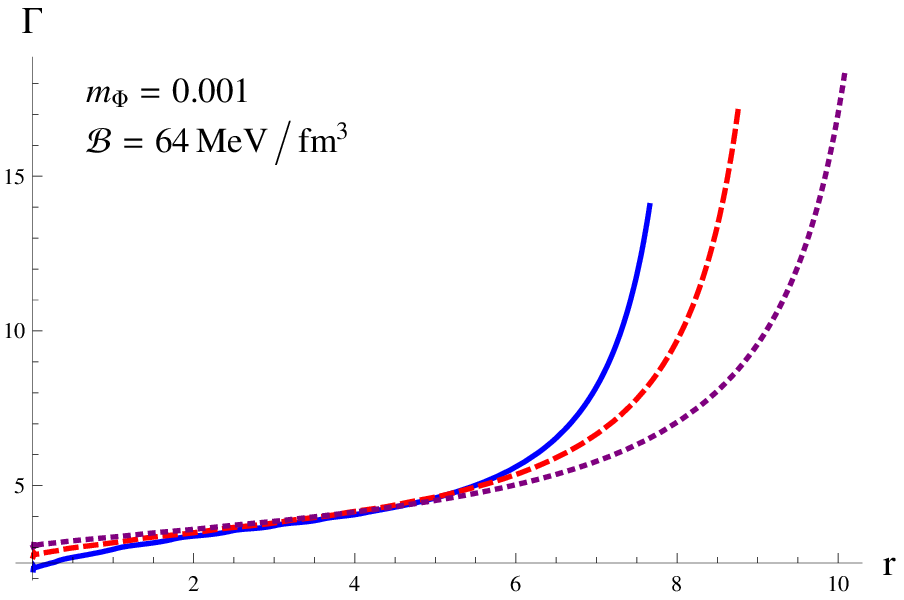,width=0.4\linewidth}\epsfig{file=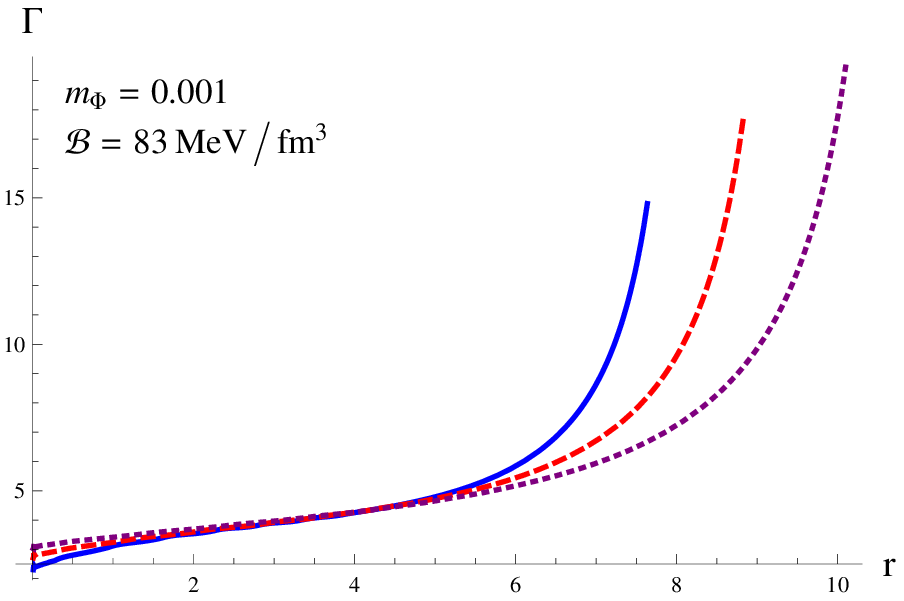,width=0.4\linewidth}\\
\epsfig{file=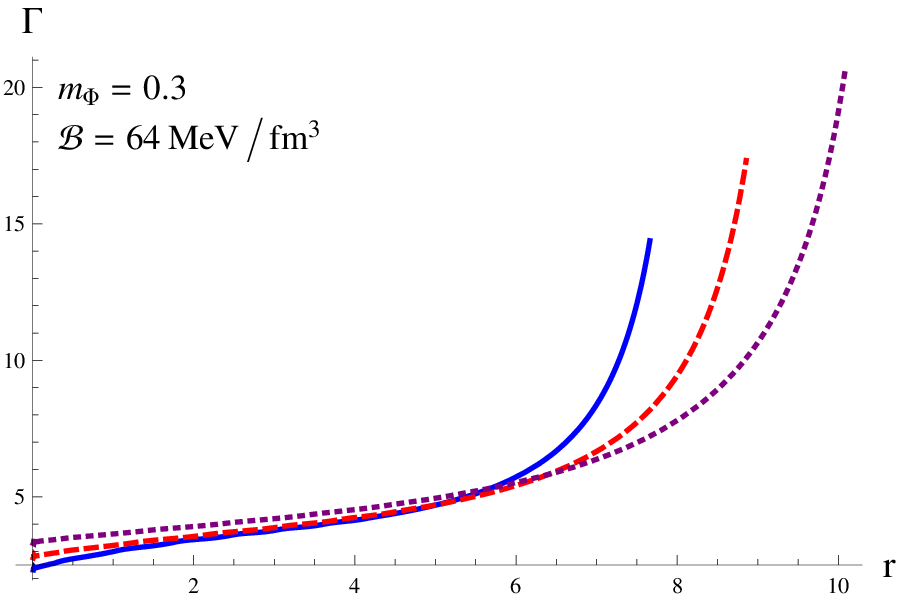,width=0.4\linewidth}\epsfig{file=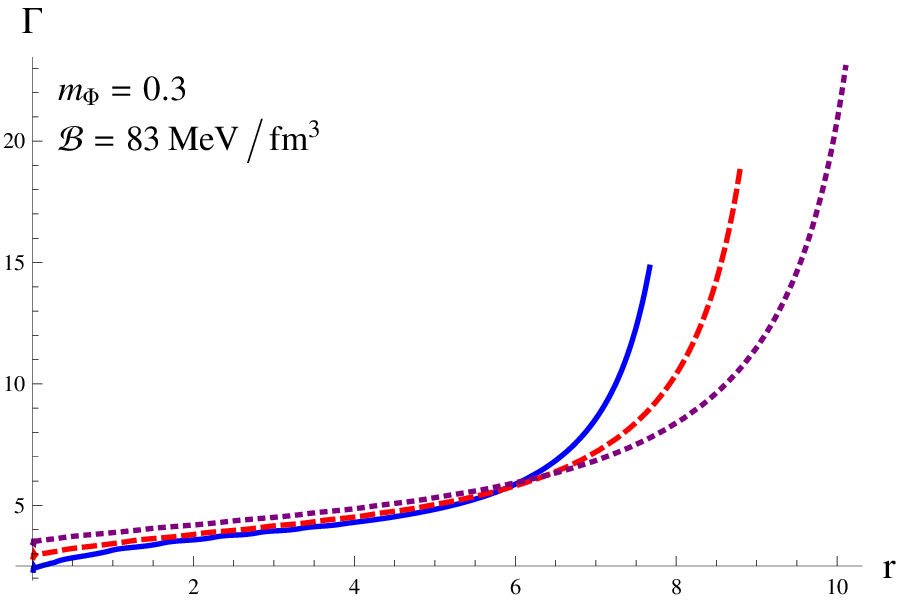,width=0.4\linewidth}
\caption{Plots of adiabatic index versus $r$.}
\end{figure}

Another commonly used tool to examine the stability of relativistic
spherical systems is adiabatic index. This indicates stiffness of
the EoS for a specific energy density by connecting the EoS with the
internal structure of the sphere. Chandrasekhar \cite{40} studied
the dynamical stability of relativistic stars against infinitesimal
radial adiabatic perturbation. Heintzmann and Hillebrandt \cite{41}
found that an anisotropic compact object will achieve stability if
the adiabatic index is greater than $\frac{4}{3}$ everywhere inside
the configuration. The expression for adiabatic index for our system
is given by
\begin{equation*}
\Gamma=\frac{p_r+\rho}{p_r}\frac{dp_r}{d\rho}
=\frac{p_r+\rho}{p_r}v_r^2.
\end{equation*}
The graphical analysis of adiabatic index in the presence of a
scalar field can be seen in Figure \textbf{9}. The value of this
index is more than $\frac{4}{3}$ for all stars which is in agreement
with the constraint \cite{41}. Hence, the stellar structure is
stable for the considered values of MBD parameters.

\section{Concluding Remarks}

In the field of astrophysics, the evolution of stellar models and
their remnants have been examined by many researchers. Strange quark
stars are hypothesized to emerge from the collapse of neutron stars
and are composed of three quark flavors.  These celestial bodies are
highly dense compact objects whose structures depend on central and
surface densities. This work investigates the possible existence of
anisotropic hypothetical objects in the background of MBD theory. We
have formulated the field equations in Jordan frame by selecting
$V(\Phi)=\frac{1}{2}m_{\Phi}^2\Phi^2,~m_\Phi=0.001,~0.3$ and
$\omega_{BD}=20,~25,~30$. A solution to the field equations has been
generated by assuming a well-behaved metric potential and embedding
class-one condition with MIT bag model. Utilizing the matching
conditions at the boundary of the star, the effective energy density
and effective pressure components have been expressed in terms of
mass and radius. The structure has finally been determined by
evaluating the constants $(A,~B,~C,~D)$ through the observed mass
and predicted radius of LMC X-4. The effect of bag constant on
stability and viability of the system has been checked through
various criteria by taking into account two values of $\mathcal{B}$
as $64MeV/fm^3$ and $83MeV/fm^3$.

Tables \textbf{1} and \textbf{2} indicate that predicted radius
increases while Figure \textbf{6} shows a decrease in mass as the
bag constant increases in the presence of massive scalar field. This
implies that the quark star becomes more dense with increase in the
bag constant. Moreover, higher values of scalar field mass lead to
more dense stellar systems as indicated in Tables \textbf{1} and
\textbf{2}. The behavior of physical attributes of the stellar
candidates has also been explored graphically. It is observed that
energy density and pressure components are finite at the center and
monotonically decrease towards the surface. The regular behavior of
state variables indicates that the system has no singularity. We
have established that the interior of star consists of normal matter
as all energy conditions are satisfied for the considered values of
the parameters as well as the bag constant.

The positive trend of anisotropy confirms the repelling force
required to save the star from further collapse. The graphical
analysis of redshift parameter which depicts that as the radius of
the star increases, the amount of redshift (complying with the limit
$Z<5.211$ \cite{37}) decreases. We have also calculated the
compactness factor and the mass-radius ratio which are in agreement
with the Buchdahl criterion \cite{36}. The graphical representation
of redshift and compactness factor of the stellar model shows
increasing behavior with a decrease in radius. Hence, a more dense
star exerts additional force on light leading to greater redshift.
Finally, we have checked the stability conditions for the prototype
stellar model using three approaches. All these imply stability of
the system coupled to a massive scalar field. However, plots of
$v_r$ and $v_\perp$ represent smooth behavior for increasing values
of $\omega_{BD}$ as shown in Figures \textbf{7} and \textbf{8}.
Hence, the celestial object is more stable for larger values of
coupling parameter. We conclude that the cosmic structure governed
by MIT bag model in the framework of MBD is consistent with all the
critical requirements and can be treated as a viable and stable
model. It is worthwhile to mention here that our results are
consistent with $f(R,T)$ theory \cite{28a}. All our results reduce
to GR for $\omega_{BD}\rightarrow\infty$.

The stellar model has also been constructed in GR with the help of
metric potentials in Eqs.(\ref{11}) and (\ref{11'}) to highlight the
effect of massive scalar field. It describes a star with estimated
radii of $5.54km$ and $5.55km$ corresponding to
$\mathcal{B}=64MeV/fm^3$ and $83MeV/fm^3$, respectively. Thus, GR
predicts smaller stars with increased central
$(9.498\times10^{16}gm/cm^3)$ and surface
($6.823\times10^{16}gm/cm^3$) densities as compared to the MBD
stellar model. Moreover, the relativistic model in GR has also
increased central pressure ($3.540\times10^{36}dyne/cm^2$,
$3.5640\times10^{36}dyne/cm^2$) for both values of the bag constant.

\section*{Appendix A}
\renewcommand{\theequation}{A\arabic{equation}}
\setcounter{equation}{0}

The components of $T^{\gamma\Phi}_{\delta}$ are obtained as
\begin{eqnarray}\label{1'}
T_0^{0\Phi}&=&e^{-\lambda}\left[\Phi''+\left(\frac{2}{r}-\frac{\lambda'}{2}
\right)\Phi'+\frac{\omega_{BD}}{2\Phi}\Phi'^2-e^\lambda\frac{V(\Phi)}
{2}\right],\\\label{1''}
T_1^{1\Phi}&=&e^{-\lambda}\left[\left(\frac{2}{r}+\frac{\nu'}
{2}\right)\Phi'-\frac{\omega_{BD}}{2\Phi}\Phi'^2-e^\lambda\frac{V(\Phi)}{2})\right],\\\label{1'''}
T_2^{2\Phi}&=&e^{-\lambda}\left[\Phi''+\left(\frac{1}{r}-\frac{\lambda'}
{2}+\frac{\nu'}{2}\right)\Phi'+\frac{\omega_{BD}}{2\Phi}\Phi'^2-e^\lambda\frac{V(\Phi)}{2}
\right].
\end{eqnarray}
Energy density and pressure components take the following form
\begin{eqnarray}\nonumber
\rho&=&\frac{1}{2r^2}\left\{\xi^2\Phi^2\left[2R^2(2M-R)\left(M\left(r^2-2R^2\right)+R^3\right)
\left(r\Phi'(r)+2\Phi(r)\right)\right]\right.\\\nonumber
&-&\left.\xi\left[r^2R^2\omega_{BD}(R-2M)\Phi'^2(r)
+2r\Phi(r)\left(\left(R^3-M\left(r^2+2R^2\right)\right)\Phi'(r)\right.\right.\right.\\\nonumber
&+&\left.\left.\left.rR^2
(R-2M)\Phi''(r)\right)-2\Phi^2(r)\left(2M(r-R)(r+R)+R^3\right)\right]\right.\\\label{21}
&+&r^2 V(\Phi)+\left.2\Phi(r)\right\},\\\nonumber
p_r&=&\frac{\xi}{4r^2}r^2R^2\omega_{BD}(2M-R)\Phi'^2(r)+2\Phi^2(r)\left(2M(r-R)(r+R)+R^3\right)\\\nonumber
&+&r\Phi(r)\left(\left(2M(r-R)(r+R)+R^3\right)\Phi'(r)+rR^2(2
M-R)\Phi''(r)\right)\\\label{22}
&+&\frac{\xi^2\Phi^2}{4r}(R^2(2M-R)\left(M\left(r^2-2
R^2\right)+R^3\right)\left(r\Phi'(r)+2\Phi(r)\right))-\mathcal{B},\\\nonumber
p_\perp&=&\frac{\Phi\xi}{4r^2}\left[\left(M\left(r^2-2R^2\right)+R^3\right)
\left(2\Phi(r)\left(2M(r-R)(r+R)+R^3\right)\right.\right.\\\nonumber
&+&\left.\left.3rR^2(R-2M)\Phi'(r)\right)\right]+\xi\left[3
r^2R^2\omega_{BD}(R-2M)\Phi'^2(r)+r\Phi(r)\right.\\\nonumber
&\times&\left.\left(\left(-2Mr^2+14M
R^2-7R^3\right)\Phi'(r)+3rR^2(R-2M)\Phi''(r)\right)-2\Phi^2(r)\right.\\\label{23}
&\times&\left.\left(2M\left(r^2-3
R^2\right)+3R^3\right)\right]-\mathcal{B}+\frac{\Phi(r)}{r^2},
\end{eqnarray}
where
$\xi=\frac{\Phi^{-1}(r)}{\left(2Mr^2e^{\frac{M(r-R)(r+R)}{R^2(R-2
M)}}+R^2(R-2M)\right)}$.

\end{document}